\documentclass[
 reprint,
 amsmath,amssymb,
 aps,
]{revtex4-2}

\usepackage{graphicx}
\usepackage{dcolumn}
\usepackage{bm}
\usepackage{dsfont}

\begin{document}

\preprint{APS/123-QED}

\title{Atomic spin-wave control and spin-dependent kicks with shaped subnanosecond pulses}
\author{Yizun~He$^1$}
\email{yzhe16@fudan.edu.cn}
\author{Lingjing~Ji$^1$}
\author{Yuzhuo~Wang$^1$}
\author{Liyang~Qiu$^1$}
\author{Jian~Zhao$^1$}
\author{Yudi~Ma$^1$}
\author{Xing~Huang$^1$}
\author{Saijun~Wu$^1$}%
\email{saijunwu@fudan.edu.cn}
\affiliation{%
$^1$Department of Physics, State Key Laboratory of Surface Physics and Key 
Laboratory of Micro and Nano Photonic Structures (Ministry of Education), 
Fudan University, Shanghai 200433, China}
\author{Darrick~E.~Chang$^{2,3}$}
\email{darrick.chang@icfo.eu}
\affiliation{
$^2$ICFO-Institut de Ciencies Fotoniques, The Barcelona Institute of Science 
and Technology,
08860 Castelldefels, Barcelona, Spain and\\
$^3$ICREA-Instituci\'o Catalana de Recerca i Estudis Avan\c{c}ats, 08015 
Barcelona, Spain
}%


\date{\today}

\begin{abstract}

The absorption of traveling photons resonant with electric dipole transitions of an atomic gas naturally leads to electric dipole spin wave excitations. For a number of applications, it would be highly desirable to shape and coherently control the spatial waveform of the spin waves before spontaneous emission can occur. This paper details a recently developed optical control technique to achieve this goal, where counter-propagating, shaped sub-nanosecond pulses impart sub-wavelength geometric phases to the spin waves by cyclically driving an auxiliary transition.
In particular, we apply this technique to reversibly shift the wave vector of a spin wave on the $D2$ line of laser-cooled $^{87}$Rb atoms, by driving an auxiliary $D1$ transition with shape-optimized pulses,  so as to shut off and recall superradiance on demand. We investigate a spin-dependent momentum transfer during the spin-wave control process, which leads to a transient optical force as large as  $\sim 1\hbar k$/ns, and study the limitations to the achieved $70\sim 75\%$ spin wave control efficiency by jointly characterizing the spin-wave control and matterwave acceleration. Aided by numerical modeling, we project potential future improvements of the control fidelity to the $99\%$ level when the atomic states are better prepared and by equipping a faster and more powerful pulse shaper. Our technique also enables a background-free measurement of the superradiant emission to unveil the precise scaling of the emission intensity and decay rate with optical depth. 
\end{abstract}
\maketitle



\section{\label{sec:level1}Introduction}
Spontaneous emission is typically a decoherence effect to avoid when levels in small quantum systems are chosen to encode information for e.g., quantum computation, simulation, or sensing~\cite{Quantuminformationandcomputation, Colloquiumgaugepotential,quantumsimulation, quantumsensing}. As spontaneous as it is, the information flow during the process can nevertheless be controlled between long-lived matter degrees of freedom and a pre-aligned single-mode electro-magnetic continuum~\cite{quantummemoryreview, quantummemoryreview2}. In particular, since the seminal work by Dicke in 1954 on super- and subradiant effects of light emission by ensembles of excited atoms~\cite{Dicke1954}, it is now well-known that the spatio-temporal properties of spontaneous emission are in principle dictated by collective properties of the atoms themselves. For spatially extended atomic ensembles, the timed phase correlations of the collective excitations, in the form of spin waves with phase-matched wave vector ${\bf k}$ satisfying $|{\bf k}|= \omega / c$, can direct superradiant emission into narrow solid angles~\cite{scully2006,singlePhotondu2012,superradianceAraujo16,superradianceRoof2016}. This collective enhancement forms the basis for many applications predicated upon efficient quantum atom-light interfaces~\cite{Fleischhauer2005, quantummemoryreview}. However, there has been relatively little work to address the question of what happens when $|{\bf k}| \neq \omega/c$ becomes strongly phase-mismatched from radiation. Within the context of atomic ensembles, complex phenomena can arise involving the combination of spatial disorder, multiple scattering of light, and dipole-dipole interactions between atoms~\cite{ZhuPRA2016, Kaiser2016, clScullyPRL09, Jun2016,Chang2020RG}, with much still left to be understood. Furthermore, within the emerging field of quantum optics with atomic arrays, such phase-mismatched states are predicted to be strongly subradiant. This forms the basis for exciting applications like waveguiding of light by the array~\cite{Du2015,Needham2019,Adams2016}, atomic mirrors~\cite{Bettles2016,Shahmoon2017, Immanuel2020}, and exotic states~\cite{selectiveRadiancePRXchang17,Sutherland2016} including emergent Weyl excitations~\cite{syznanov2016} and topological guided edge modes~\cite{Perczel2017,Bettles2017}, and the generation of highly correlated, ``fermionized'' states ~\cite{Molmer2019}. One major bottleneck to exploring and controlling all of these phenomena is the fact that any optical pulses used to manipulate atomic excited states, being radiation waves, most naturally excite phase-matched spin waves, while spin waves with $|{\bf k}| \neq \omega/c$ are naturally decoupled from light. What is needed then is a technique to efficiently and coherently alter the phase-matching condition of collective atomic excitations in the temporal domain --- that is, to modify the wave vector of the spin waves and coherently convert between superradiant and subradiant modes --- on rapid time scales faster than the typical emission time of atoms themselves.

Previously, in a short Letter~\cite{coSub}, we described an experimental realization of coherent dipole spin-wave control, based upon rapidly and cyclically driving an auxiliary transition with counter-propagating control pulses to robustly imprint spin- and spatially-dependent geometric phases onto the atoms. The resulting ${\bf k}$-space shifts lead to spin waves in a $^{87}$Rb gas with strongly mismatched wave vectors $|{\bf k}| \neq \omega/c$ off the light cone and suppressed superradiant emission. Later, the mismatched spin waves are shifted back onto the light cone to cooperatively emit again on demand. In this paper, we expand upon key details that were previously omitted, describing in greater detail the experimental implementation and theoretical modeling of the geometric control technique, and additional research advances enabled by the technique. Furthermore, we carefully characterize the fidelity of our control sequence, and discuss how current limitations can be overcome. 


Beyond a detailed discussion of our spin-wave control technique, here, we also give an example of its use towards the precise study of fundamental quantum optical phenomena. In particular, we exploit the ability to shift the dipole spin-wave vector to measure phase-matched forward collective emission in a background-free manner. The forward collective emission has been studied previously as a strong signature of cooperative enhancement of light-atom interactions in cold atomic gases~\cite{superradianceRoof2016,superradianceAraujo16, Jun2016}.  One challenge to measure the forward cooperative emission in these experiments is related to the fact that the exciting beam, typically with a much stronger intensity, is in the same direction as the forward emission and contributes a large background. Here, the ability to shift the spin wave vector in our case allows us to detect the forward superradiant emission from a different, background-free direction. We experimentally quantify the scaling of the emission intensity $i_N\propto N^2$ and collective decay rate $\Gamma_N/\Gamma=(1+\overline{{\rm OD}}/4)$~\cite{clScullyPRL09,Kaiser2016, Sutherland2016,ZhuPRA2016}, for superradiant emission involving $N$ atoms with an average optical depth $\overline{{\rm OD}}$.

This paper also brings together two seemingly unrelated phenomena: the control of 
collective dipole radiation, and the acceleration of the free emitters. Accompanying the nanosecond spin-wave ${\bf k}$-vector shifts, we observe a strong spin-dependent optical force that accelerates the atomic sample at a $\sim 10^7$~m/s$^2$ transient rate. Similar techniques of cyclic rapid 
adiabatic passage have been studied in pioneering work by Metcalf {\it et al.}~\cite{accMetcalf2007,nonAdiabaticMetcalf2007} as a robust way to impart strong optical forces to neutral atoms and molecules, with important applications in laser cooling~\cite{slowCampbell2014, MetcalfRevew2017}. The $\sim 1\hbar k$ per nanosecond optical force in this work is among the highest~\cite{MetcalfRevew2017, campbell2019}. The negative impacts by the spin-dependent acceleration to the spin-wave control is negligible in this work, and can be mitigated with lattice confinements in future experiments~\cite{coSub}. On the other hand, the combined effects may open interesting opportunities at the interface of quantum optics and atom interferometry~\cite{DaWei2014superAI}.

The remainder of the paper is structured as follows. First, in Sec. II, we provide a simple, idealized theoretical description of our protocol to manipulate spin waves by rapid geometric phase patterning using shaped sub-nanosecond pulses. In Sec. III, we detail the experimental implementation of the coherent control and discuss the background-free detection of the cooperative emission as well as matter-wave acceleration effects. Here we also generalize the simpler theoretical discussion for quantifying the efficiency of our protocol in the face of various imperfections. We summarize this work in Sec. IV, and discuss methods and technology for improvements of the spin-wave control efficiency to the $\sim$99\% level. To ensure completeness and to provide better context, key ideas from Ref.~\cite{coSub} are repeated in  this work.

\begin{figure*}
    \includegraphics[width=0.90\textwidth]{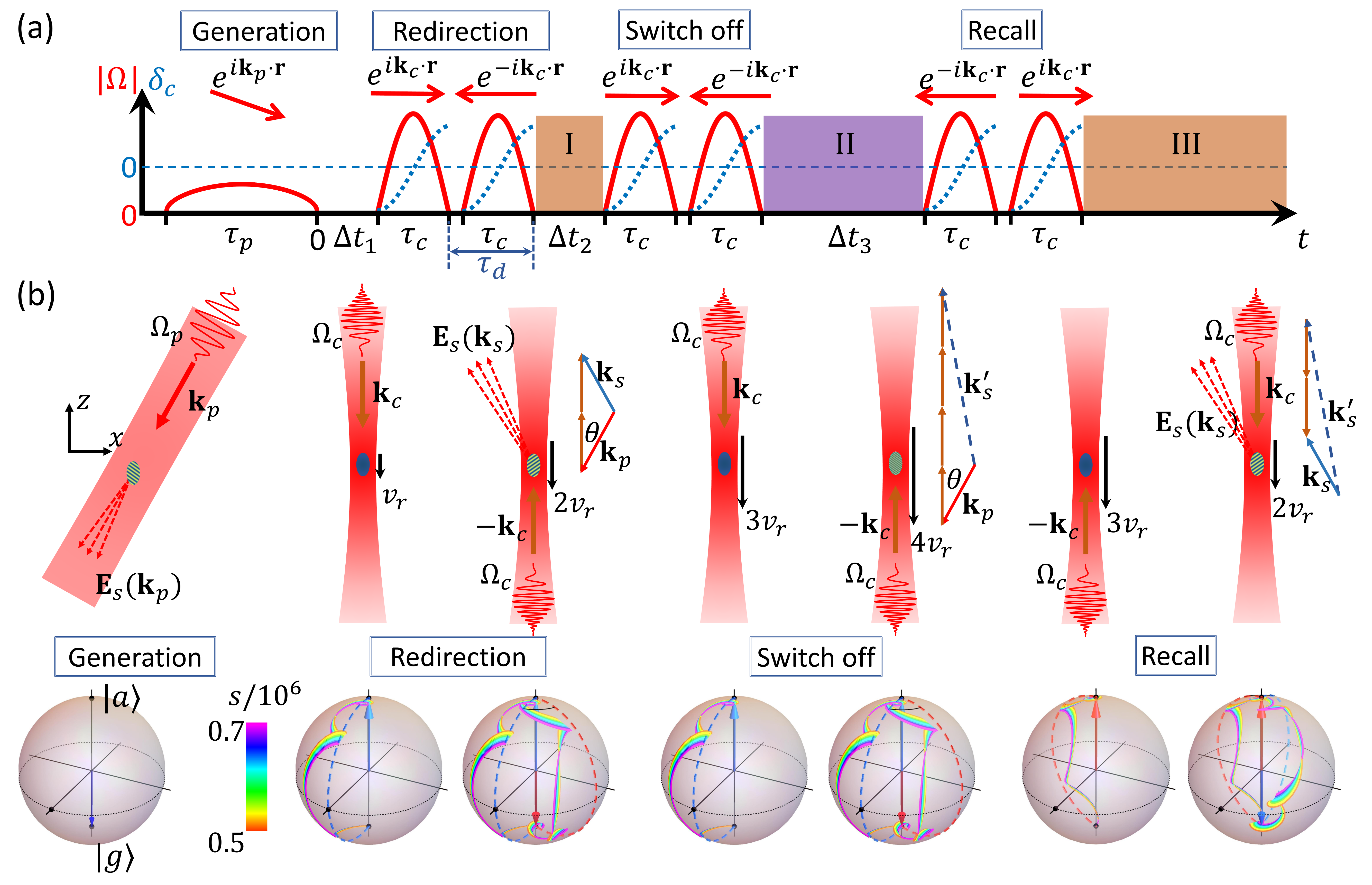}
    \caption{Schematic of the experiment to demonstrate error-resilient optical 
    spin-wave control. (a) Schematic timing sequence for the amplitudes of the 
    probe and control pulse Rabi frequency $|\Omega|$ (red solid lines), and the 
    instantaneous detuning of the control pulse $\delta_c$ (blue dashed lines) 
    from the $|g\rangle -|a\rangle$ transition. (b) Top (from left to right): 
    Generation and control of optical spin wave with the probe beam followed by the control pulse-induced redirection, switch off and recall of the collective spontaneous emission. The  $|g\rangle-|e\rangle$ electric dipole spin wave is illustrated with fringes 
    in the atomic sample. Each optical control also imparts a spin-dependent kick, 
    which leads to momentum transfer with $v_r=\hbar k_c/m\approx 5.8$~mm/s, 
    $m$ being the atomic mass of $^{87}$Rb. The drawings are not to actual scales, in 
    particular, the phase-matching angle 
    $\theta=\arccos\frac{\lambda_{D2}}{\lambda_{D1}}\sim 11.1^{\circ}$ is exaggerated for 
    clarity. Bottom: Bloch-sphere 
    representation of the projected $|g\rangle-|a\rangle$ state dynamics for an 
    atom at a representative position $\bf r$. An ensemble of trajectories with different control pulse peak intensity parameters $s$ is displayed. The quasi-adiabatic control 
    ensures that the geometric phase writing is insensitive to small deviations of $s$ from 
    $s\sim 0.6\times 10^6$, for $\tau_c \Gamma_{D1} =0.02\sim 0.03$ in this 
    paper.}\label{figTimingsequence}
\end{figure*}

\section{Principles}

\subsection{Preparation and control of optical spin waves\label{secSimple}}

Single collective excitations of an atomic ensemble, consisting of $N$ atoms 
with ground state $|g\rangle$ and excited state $|e\rangle$, are naturally 
described by ``spin-wave'' excitations or timed-Dicke states of the form 
$|\psi_{\rm TD}({\bf k}_p)\rangle=S^+({\bf k}_p)|g_1,g_2,\cdots,g_N\rangle$. 
Here $S^+({\bf k}_p)=\frac{1}{\sqrt{N}}\sum_i{e^{i {\bf k}_p\cdot {\bf r}_i} 
|e_i\rangle\langle g_i|}$ denotes a collective spin raising operator and 
${\bf k}_p$ the corresponding wave vector. For example, a weak coherent state involving such excitations is naturally generated by an incoming plane-wave ``probe pulse'' with Rabi frequency $\Omega_p$ and duration $\tau_p\ll 1/\Omega_p$, if 
$\tau_p$ is in addition short enough that light re-scattering effects are negligible. As  such, the magnitude of the wave vector $|{\bf k}_p|=\omega_{eg}/c$, with $\omega_{eg}$ being the atomic resonance frequency, matches that of free-space radiation. As is well-known~\cite{scully2006,superradianceAraujo16, superradianceRoof2016}, such ``phase-matched'' spin wave excitations in a large sample with size $\sigma\gg 1/k_p$ and $N>k_p^2 \sigma^2$ radiate efficiently and in a collectively enhanced fashion, much like a phased array antenna, into a small solid angle $\theta^2$ with $\theta\sim 1/k_p\sigma$ around the forward ${\bf k}_p$ direction.



We are interested in spin waves with strongly mismatched $|{\bf k}|\neq \omega_{eg}/c$ ``off the light cone'', in which case the cooperative emission is prohibited. In an ensemble with random atomic positions $\{{\bf r}_i\}$, as in this paper, such a spin wave is expected to decay with a rate near the natural linewidth~\cite{scully2015}, as the fields emitted by different atoms in any direction average to zero, but with non-zero variance. On the other hand, in an ordered array of atoms, the destructive interference in all directions can be nearly perfect, leading to strong subradiance that forms the basis of the exciting applications mentioned earlier~\cite{Du2015, Needham2019, Adams2016, Bettles2016, Shahmoon2017, Sutherland2016, selectiveRadiancePRXchang17, syznanov2016, Perczel2017, Bettles2017,Immanuel2020}. It is therefore highly compelling to have a technique to shift the spin waves in ${\bf k}$ space on a time scale much faster than the spontaneous emission lifetime. To achieve the required ${\bf k}$-vector shift, we consider a unitary transform $U_c(\Delta {\bf k})$ such that $S^{+}({\bf k}+\Delta {\bf 
k})=U_c(\Delta {\bf k})S^{+}({\bf k})U_c^{\dagger}(\Delta {\bf k})$. The required control is a class of state-dependent phase-patterning operations, which can be decomposed into spatially dependent phase gates for each two-level atom as (with $\sigma_{z,i}=|e_i\rangle\langle 
e_i|-|g_i\rangle\langle g_i|$)

\begin{equation}
  U_c(\Delta {\bf k})=\prod_i^N e^{\frac{i}{2}\Delta {\bf k}\cdot {\bf r}_i 
\sigma_{z,i}}.\label{equU0z}
\end{equation}

While we have thus far focused on the manipulation of spin-wave excitations, by considering the position ${\bf r}_i$ as an operator as well, one sees that the transformation of Eq.~(\ref{equU0z}) also imparts opposite momentum boosts to the $g$, $e$ components of a freely moving atom. Techniques for realizing such 
spin-dependent kicks have been well-developed in the community of 
atom interferometry and ion-based quantum information processing typically on 
Raman transitions~\cite{Monroe2014,Monroe2017,houge2018}. Here, we demonstrate a high-efficiency process based upon rapid manipulation of a strong optical transition for the spin-wave control.


\subsection{Error-resilient spin-wave control\label{SecERControl}}

To implement $U_c(\Delta {\bf k})$ in Eq.~(\ref{equU0z}) on a strong dipole 
transition in a large atomic sample, we combine the geometric phase 
method suggested in Ref.~\cite{scully2015} with optical rapid adiabatic passage techniques as 
in Refs.~\cite{accMetcalf2007,nonAdiabaticMetcalf2007} for the necessary control speed, 
precision, and intensity-error resilience. In particular, we consider two nearly identical  optical ``control'' pulses with 
Rabi frequency $\Omega_c(t)e^{i \varphi_{1,2}}$ and instantaneous detuning 
$\delta_c(t)$ to drive the auxiliary $|g\rangle\rightarrow|a\rangle$  and 
$|a\rangle\rightarrow|g\rangle$  transitions, respectively. By optical pulse 
shaping, the cyclic transition can be driven with high precision in a manner that is largely insensitive to laser intensity. We adapt a simple choice of 
$\Omega_c(t)=\Omega_0\sin(\pi t/\tau_c)$ and $\delta_c(t)=-\delta_0\cos(\pi 
t/\tau_c)$ to achieve quasi-adiabatic population inversions at optimal 
$\{\Omega_0,\delta_0\}$ within the $\tau_c$ pulse duration~\cite{accMetcalf2007,nonAdiabaticMetcalf2007}. To connect to 
the experimental setup (Figs.~\ref{figTimingsequence} and \ref{figTwodelayline}), after preparing the $S^+({\bf k}_p)$ spin wave excitation, we send the first control pulse to the atomic sample, propagating along the direction given by wave vector $\pm{\bf k}_c$, driving $|g\rangle\rightarrow|a\rangle$. Subsequently, a second control pulse with $\mp{\bf k}_c$ interacts with the sample after a $\tau_d$ delay, driving $|a\rangle\rightarrow|g\rangle$. Although every ground-state atom in the original $|g\rangle-|e\rangle$ spin wave returns to $|g\rangle$, the area enclosed around the Bloch sphere by the state vector causes each atom to pick up a spatially dependent geometric phase $\varphi_{\rm G}({\bf r})=\pi+\Delta {\bf k}\cdot {\bf r}_i$, with $\Delta {\bf k}=\pm 2{\bf k}_c$ to fully exploit the resolution of the optical phase. The ideal state-dependent phase patterning achievable in the short $\tau_{c,d}$ 
limit can be formally expressed within the $\{|g\rangle,|e\rangle\}$ space as
\begin{equation}
U_g(\Delta {\bf k})=\prod_i^N{\left ( |e_i\rangle\langle e_i|- e^{i \Delta {\bf k}\cdot {\bf r}_i} |g_i\rangle\langle g_i| \right )},\label{equUc}
\end{equation}
which performs a ${\bf k}\rightarrow {\bf k}-\Delta {\bf k}$ shift to spin-wave excitations the same way as $U_c(-\Delta {\bf k})$ in Eq.~(\ref{equU0z}).

Although the phase patterning operation in Eq.~(\ref{equUc}) could in principle be achieved without the control pulse shaping, practically, the control laser intensity inhomogeneity across the large sample would translate directly into spin control error, as in most nonlinear spectroscopy experiments~\cite{cundiff2013, Fuller2015, Oliver2018}, leading to degraded average control fidelity if a highly uniform laser intensity profile cannot be maintained. To fully exploit the intensity-error resilience~\cite{geometricAspect2012} offered by the SU(2) geometry of the two-level control, the optical pulses need to be shaped on a time scale fast enough to suppress the spontaneous emission, and also slow enough to avoid uncontrolled multi-photon couplings in real atoms. We note that optical methods for two-level rapid adiabatic passage~\cite{Loy1974} itself are well-developed for population transfers in atoms and molecules using ultrafast lasers~\cite{GoswamiPR2003,Shapiro2008}. However, these ultrafast techniques typically demand control field Rabi frequencies $\Omega_c$ beyond the THz level with intense pulses at low repetition rates, not easily adaptable to our desired goals. The strong fields may also cause non-negligible multi-level couplings or even photo-ionization beyond the desired multiple population inversions. Instead of using ultrafast lasers, here we develop a wide-band pulse-shaping technique based on fiber-based sideband electro-optical modulation of a cw laser~\cite{coSub}, with up to 13~GHz modulation bandwidth, to support the flexibly programmable error-resilient spin wave control. Compared with previous work on spectroscopy based upon perturbative nonlinear optical effects~\cite{cundiff2013, Fuller2015, Oliver2018}, our technique is unique in that we steer the atomic state over the entire Bloch sphere of the two-level system to achieve the geometric robustness toward perfect spin-wave control set by Eq.~(\ref{equUc}).

\subsection{Dynamics of controlled emission\label{secSimpleDynamics}}

Here, we go beyond Ref.~\cite{coSub}, to discuss the expected emission characteristics of phase-matched and mismatched spin waves, which will constitute one of the key observables to verify our coherent control technique and quantify its efficiency. Later, in Sec.~\ref{secDensity}, we will experimentally verify the predicted optical depth dependence for the phase-matched case. To specifically relate to experimental control of a laser-cooled gas in this 
work, we consider an atomic sample with a smooth profile $\varrho({\bf 
r})=\langle \sum_i \delta({\bf r}-{\bf r}_i) \rangle$ nearly spherical with 
size $\sigma\gg 1/|{\bf k}_p|$ and at a moderate density with $\varrho<|{\bf 
k}_p|^{3}$. Formally, the quantum field emitted by a collection of atoms on the $|e\rangle-|g\rangle$ transition can be expressed in terms of the atomic spin coherences themselves, $\hat {\bf E}_{s}({\bf r})=\omega_{eg}^2 
/\varepsilon_0 c^2\sum_i {\bf G}({\bf r}-{\bf r}_i,\omega_{e g}) \cdot {\bf d}_{e g} 
|g_i\rangle\langle e_i|$~\cite{selectiveRadiancePRXchang17}. Here ${\bf G}({\bf r},\omega_{eg})$ is 
the free-space Green's tensor of the electric field, which physically describes the field at position $\bf r$ produced by an oscillating dipole at the origin, and ${\bf d}_{eg}=d_{eg}{\bf e}_d$ is 
the transition dipole moment. For the single-excitation timed Dicke state, one can define a single-photon wave function ${\bf \boldsymbol{\varepsilon}}_{\bf k}({\bf r})=\langle g_1, 
g_2,...,g_N|\hat {\bf E}_s |\psi_{\rm TD}({\bf k}) \rangle$, which describes the spatial profile of the emitted photon. Of particular interest will be the field emitted along the direction ${\bf k}$ at the end of the nearly spherical sample, and at a transverse position ${\bf r}_\perp$. Within the so-called ``Raman-Nath'' regime where diffraction is negligible, we obtain 
$\overline{{\bf \boldsymbol{\varepsilon}}_{\bf k}}({\bf r})$ averaged over random $\{{\bf 
r}_i\}$ as (also see Appendix~B)
\begin{equation}
  \overline{\boldsymbol{\varepsilon}_{\bf k}}({\bf r})\approx {\bf e}_d \sqrt{\frac{ \hbar 
\omega_{eg}}{8 \varepsilon_0 c}}\sqrt{N \sigma_r \Gamma}\varrho_c({\bf 
r}_{\perp},\delta k) e^{i \omega_{e g} r_{\bf k}/c }.\label{equEs}
\end{equation}
Here $r_{\bf k}={\bf r}\cdot {\bf k}/|{\bf k}|$, and $\varrho_c({\bf r}_{\perp},\delta 
k)=\frac{1}{N}\int \varrho({\bf r}) e^{i\delta k r_{\bf k}} d r_{\bf k}$ is 
a generalized (and normalized) column density, where in general we allow for a wavenumber $\delta k=|{\bf k}|-\omega_{e g}/c$ that is mismatched from radiation. 
$\sigma_r=\omega_{eg} \alpha_i/c$ is the resonant absorption cross-section, 
and the imaginary part of the resonant polarizability $\alpha_i$ is related 
to the dipole element ${\bf d}_{e g}$ and $\Gamma$ through $ \hbar \Gamma \alpha_i =2 |{\bf d}_{e g}|^2$. While ${\bf d}_{eg}$ and $\Gamma$ are directly related for two-level atoms, this formula also generalizes to atoms with level-degeneracy.

Equation~(\ref{equEs}) describes the possibility of both enhanced or suppressed collective 
emission associated with the spin-wave excitation at a $\delta k$ 
radiation wave-number mismatch. A well-known consequence of Eq.~(\ref{equEs}) 
is that when the spin-wave excitation is by a weak probe pulse with 
wave vector ${\bf k}_p$ (see Fig.~\ref{figTimingsequence}), the atoms act as a phased 
antenna array with $\delta k=0$, and light in the forward direction along 
${\bf k}_p$ is re-emitted at an enhanced rate within an angle $\theta\sim 
1/(|{\bf k}_p| \sigma)$~\cite{scullyscience09}. As our probe pulse is in a weak coherent state, the 
timed Dicke state is excited with a population of $N\theta_p^2$, where 
$\theta_p=\frac{1}{2}\int \Omega_{p} dt$ is the time-integrated Rabi 
frequency of the probe pulse ($\theta_p\ll 1$). Assuming that the spatial 
profile of Eq.~(\ref{equEs}) does not change significantly during the 
emission process, one can integrate the intensity of light predicted by 
Eq.~(\ref{equEs}) over space, and arrive at the following time-dependent 
collective spontaneous emission rate:

\begin{equation}
i_{{\bf k}_p}(t)\approx N \theta_p^2 (\overline{{\rm OD}}_p/4) \Gamma 
e^{-(1+\overline{{\rm OD}}_p/4)\Gamma t}.\label{equi2}
\end{equation}
Here $\overline{{\rm OD}}_p\equiv \int {\rm OD}_p^2({\bf r}_{\perp}) /\int {\rm OD}_p({\bf 
r}_{\perp})\propto N$ is the average optical depth, and ${\rm OD}_p({\bf 
r}_{\perp})=N \varrho_c({\bf r}_{\perp},\delta k=0)\sigma_r$.
The exponential factors of $e^{-\Gamma t}$ and $e^{-\overline{{\rm OD}}_p\Gamma 
t/4}$ account for the (non-collective) decay into $4\pi$ and enhanced 
emission along the phase-matched ${\bf k}_p$ direction, respectively.

We now consider what happens if, immediately following the probe pulse at 
$t=0$, we apply the ideal spin-wave control, which imprints a geometric phase 
and transforms the original timed-Dicke state along ${\bf k}_p$ according to 
Eq.~(\ref{equUc}) (finite delay times and other imperfections can be 
straightforwardly included, as discussed in Sec. III). As the original state 
has a well-defined wave vector, the application of Eq.~(\ref{equUc}) simply 
creates a new timed Dicke state with new wave vector ${\bf k}_s={\bf 
k}_p-2{\bf k}_c$. Then, two distinct cases emerge. The first is that $|{\bf 
k}_s|\approx \omega_{eg}/c$, in which case the spin wave is again 
phase matched to radiation, and an enhanced emission rate like 
Eq.~(\ref{equi2}) is again observed, but with the majority of emission 
``redirected'' along the new direction ${\bf k}_s$. The second, and more 
intriguing, possibility is that $|{\bf k}_s|$ is significantly mismatched 
from $\omega_{eg}/c$. In that case, there is no direction along which 
emission can be constructively and collectively enhanced. For the case of 
our disordered ensemble, this results in the emission rate into the same 
solid angle in absence of phase-matching as:
\begin{equation}
    i_{{\bf k}_p}'(t)\approx N \theta_p^2/(|{\bf k}_p|^2\sigma^2) \Gamma e^{-\Gamma 
t},\label{equi3}
\end{equation}
i.e., the emission reduces to an incoherent sum of those from single, 
isolated atoms. This is due to the random positions of the atoms, such that 
the field of emitted light in any direction tends to average to zero, 
but with a non-zero variance [corrections to Eq.~(\ref{equi3}) are expected at high densities due to  granularity of the atomic distribution, which make the problem quite complex in general~\cite{ZhuPRA2016,shortPaper2,Chang2020RG}.]. However, in an ordered array of atoms, the 
destructive interference in all directions can be nearly perfect, leading to 
a decay rate much smaller than $\Gamma$. The ability to generate excited 
states with extremely long lifetimes is key to all of the applications 
mentioned in the Introduction~\cite{Du2015, Needham2019, Adams2016,Immanuel2020, Bettles2016, Shahmoon2017, selectiveRadiancePRXchang17, syznanov2016, Perczel2017, Bettles2017,DEC2019PRA}.

\section{Experimental Results}

\begin{figure*}
    \includegraphics[width=0.90\textwidth]{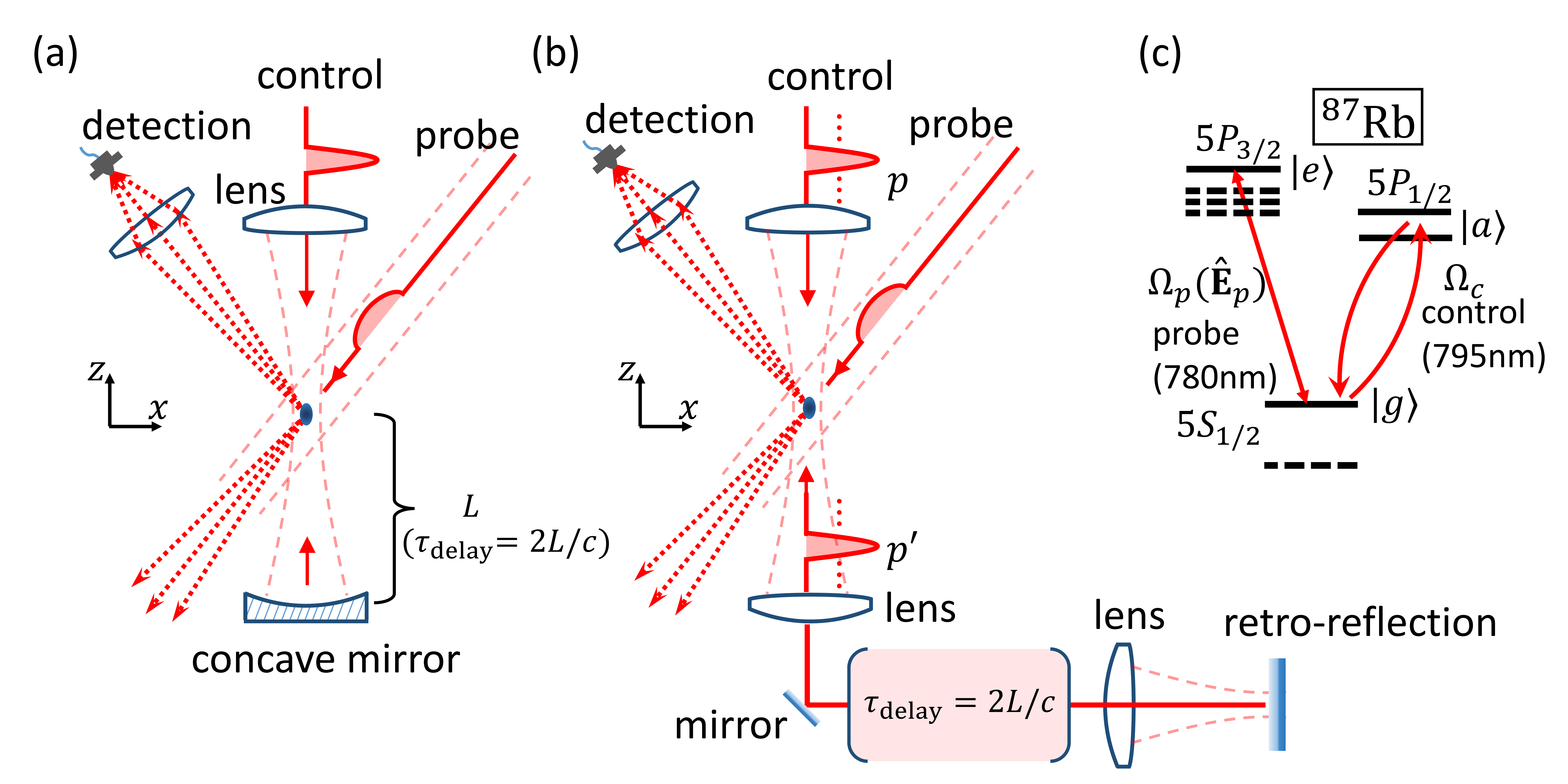}
    \caption{Setup schematics for spin-wave control based on optically delayed retro-reflections. The schematics share the same relative orientations of the laser beams as in Fig.~\ref{figTimingsequence}. In (a), the incoming pulse is retro-reflected with a delay $\tau_{\rm delay} = 1.36$~ns for $L\approx 200$~mm to interact twice with the atomic sample. In (b), the pre-programmed pulse (marked with $p'$) is stored in the much longer optical delay line with $L\approx 15$~m and $\tau_{\rm delay} \approx 100$~ns, timed to collide with a new incoming pulse (marked with $p$). (c) The $D1$ and $D2$ level diagram for the $^{87}$Rb with hyperfine structures, and the laser coupling scheme in this paper.}\label{figTwodelayline}
\end{figure*}

\subsection{Experimental methods\label{secResult}}

In this work the dipole spin wave excitation is implemented on the 
$^{87}$Rb $5S_{1/2}-5P_{3/2}$ $D2$ line between hyperfine ground state 
5$S_{1/2}$~$F=2$ and excited state 5$P_{3/2}$~$F'=3$, represented by 
$|g\rangle$ and $|e\rangle$ in Fig.~\ref{figTwodelayline}, respectively. The 
transition wavelength is $\lambda_{\rm D 2}=780$~nm (with $k_p=2\pi/\lambda_{\rm 
D 2}$), and the natural linewidth is $\Gamma_{\rm D 2}=2\pi\times 6.07$~MHz. 
We prepare $N\sim 10^4$ $^{87}$Rb atoms in $|g\rangle$ in an optical dipole 
trap with up to $\sim 5\times 10^{12}$/cm$^3$ peak density and $T\sim 
20~\mu$K temperature. After the atoms are released from the trap, the 
dipole excitation is induced by a $\tau_p=3 \sim 5$~ns, $I_p \approx 10~$mW/cm$^2$ 
resonant $D2$ probe pulse. The Gaussian probe beam has a $w_p\approx 50~\mu$m 
waist, which is much larger than the $1/e$ radius of atomic density profile 
$\sigma\approx 7~\mu$m, validating the plane-wave excitation picture.

The auxiliary transition is implemented on the $D1$ line between $|g\rangle$ 
and $|a\rangle$, with $|a\rangle$ representing the 5$P_{1/2}$~$F'=1,2$ 
levels, with $\lambda_{D1}=795$~nm, $k_c=2\pi/\lambda_{D1}$ and 
$\Gamma_{D1}=2\pi\times 5.75$~MHz. The spin wave $U_g$ control as in 
Eq.~(\ref{equUc}) is implemented by cyclically driving the $D1$ transition 
with the counter-propagating chirped pulses, with Rabi frequency 
$\Omega_c(t)=\Omega_0 \sin(\pi t/\tau_c)$ and instantaneous detuning 
$\delta_c(t)=-\delta_0 \cos(\pi t/\tau_c)$ (defined relative to the midpoint of the 
5$S_{1/2} F=2$ $-$ 5$P_{1/2} F'=1,2$ hyperfine lines), so as to phase-pattern the 5$S_{1/2} F=2$ 
atoms without perturbing the $5P_{3/2}$ level due to the large $D1$-$D2$ 
transition frequency difference.
Utilizing our sub-nanosecond pulse shaping technology as detailed in Ref.~\cite{coSub}, with $\sim$20~mW of peak power, peak intensity parameter $s\sim 10^6$ 
($s\equiv I/I_{s1}$ and $I_{s1}= 4.49~$mW/cm$^2$ is the $D1$ transition 
saturation intensity~\cite{Steck2015}) and peak Rabi frequency 
$\Omega_0=\sqrt{s/2}\Gamma_{D1}$ at GHz level are reached by focusing 
the ${\bf k}_c$-control beam into a waist of $w\approx 13~\mu$m at the 
atomic sample.

We choose the polarization for the probe and control lasers to be along 
${\bf e}_y$ and ${\bf e}_x$, respectively. Taking the ${\bf e}_x-$direction 
as the quantization axis, the $\pi$-control couplings to $5P_{1/2}$ would 
be with equal strengths and detunings for all five 5$S_{1/2} F=2, m_F$ 
Zeeman sub-levels, and with vanishing hyperfine Raman coupling, if the 
5$P_{1/2}$ hyperfine splitting $\Delta_{D1, {\rm hfse}}=2\pi\times 814.5$~MHz 
can be ignored. The approximation helps us to establish the simple two-level 
control picture in Fig.~\ref{figTwodelayline}(c), even for the real atom. Practically, 
the hyperfine dephasing effects can be suppressed for atoms in the $|m_F|=0,2$ 
Zeeman sublevels by adjusting the optical delay $\tau_d$  to match 
$2\pi/\Delta_{D1, {\rm hfse}}\approx 1.23~$ns. The hyperfine effect  
more severely impacts the $|m_F|=1$ atoms through both intensity-dependent dephasing 
and non-adiabatic population losses. These hyperfine effects are 
suppressible with better Zeeman-state preparations, or by faster controls 
with $\tau_{c}\Delta_{D1, {\rm hfse}}\ll1$ while setting $\tau_d$ to be a multiple of $2\pi/\Delta_{D1, {\rm hfse}}$.

To experimentally implement the spin-wave $U_g(\Delta {\bf k})$ control as in Eq.~(\ref{equU0z}) with $\Delta {\bf k}=\pm 2 {\bf k}_c$, counter-propagating pulses are sent to the atomic sample using retro-reflection with optical delay lines [Figs.~\ref{figTwodelayline}(a) and \ref{figTwodelayline}(b)]. The ability to control the sign is important for the turn-off and recall of superradiance. Going beyond Ref.~\cite{coSub}, we discuss how to implement this using two types of optical delays. In the first type [Fig.~\ref{figTwodelayline}(a)], an incoming pulse and its retro-reflection by a $R=200$~mm concave mirror at a $L\approx 200$~mm distance outside the vacuum interact twice with the sample with a $\tau_{\rm delay} =2L/c= 1.36$~ns relative delay. This design conveniently enables the $|g\rangle \rightarrow |a\rangle \rightarrow |g\rangle$ cyclic transition driven by a ${\bf k}_c$ and then a $-{\bf k}_c$ pulse, thereby accomplishing a $U_g(2 {\bf k}_c)$ operation with nearly identical pulse pairs. However, the ordered arrivals for the $\pm{\bf k}_c$ pulses rule out the possibility of realizing efficient $U_g(-2{\bf k}_c)$ control. 

To reverse the time order for the  $\pm {\bf k}_c$ pulses, we introduce a second type of optical delay  as in Fig.~\ref{figTwodelayline}(b). Here the delay distance $L\approx 15$~m and the associated delay time $\tau_{\rm delay} \approx 100$~ns are long enough that we are able to temporally store a few pre-programmed pulses on the side of the atomic sample opposite to the incoming direction. In particular, before the preparation of the $|g\rangle-|e\rangle$ spin waves, up to three pre-programmed $D1$ control pulses, such as the one marked with ``$p'$'' in Fig.~\ref{figTwodelayline}(b), are initially sent to the optical delay line. After all these pulses pass through the atomic sample, the atoms are excited and then decay into the $F=$1, 2 ground states within the delay time $\tau_{\rm delay} \gg 1/\Gamma_{D1}$.  At the moment when these stored pulses are coming back, additional control pulses [such as the one marked with ``$p$'' in Fig.~\ref{figTwodelayline}(b)] with readjusted pulse properties are sent to collide with the pre-programmed pulses to form the control sequence including spin-wave controls $U_g(\pm 2{\bf k}_c)$ in a designed order [Fig.~\ref{figTimingsequence}(a)].
The second optical delay method is therefore more flexible for parametrizing the control pulse pairs to reversibly shift the spin waves in ${\bf k}$-space on demand. We notice that the first pass of the stored optical pulses causes some atom losses (to the $F=1$ ground states which are dark to the following spin-wave excitations). The amount of loss is a function of the (unwanted) $D1$ excitation efficiency, and is therefore correlated to the pulse number, relative delays and shapes. By combining proper timing of the stored pulses with numerical modelings, the loss effects can be controlled in measurements where the atom numbers are important~\cite{shortPaper2}. In addition, in this work the extra beam steering optics to fold the delay line onto the optical table introduces extra optical power loss (30\%) to the retro-reflected pulses, challenging the range of intensity resilience for the control operation (also see Figs.~\ref{figTimingsequence}~and~\ref{figAcc}). Therefore, to optimally operate the control with the second mode of optical delay, we typically readjust the pulse-shaping parameters and in particular to balance the power of the pulse pairs. To precisely measure the optical delay time $\tau_{\rm delay}$, a single $D1$ control chirped pulse is sent to the optical path to efficiently excite the atomic sample twice. The two fluorescence signals separated by $\tau_{\rm delay}$ are collected to precisely measure $\tau_{\rm delay}$ with $\sim0.1$~ns accuracy.

Additional details on the experimental measurements are given in Appendix A. With the experimental methods, we can precisely perform multiple $D1$ controls to shift the dipole spin-wave excitation $S^{+}({\bf k})$ on the $D2$ transition, from the original value of ${\bf k}={\bf k}_p$ to the new wave vectors ${\bf k}_p-2 n {\bf k}_c$ (with $n=1,2$ in this paper) in a reversible manner. To benchmark the control quality, we set up the photon detection path along a finely aligned ${\bf k}_s={\bf k}_p-2 {\bf k}_c$ direction to meet the $|{\bf k}_s|=\omega_{eg}/c$ 
redirected phase-matching condition (Fig.~\ref{figTimingsequence}). As such, following the ${\bf k}_p$ spin wave preparations a ${\bf k}\rightarrow {\bf k}-2{\bf k}_c$ shift can redirect the forward superradiance to ${\bf k}_s$ for its background-free detection. We collect the 
${\bf k}_s$-mode superradiance with a NA=0.04 objective for detection by a multi-mode fiber coupled single photon 
counter. To enhance the measurement accuracy, an optical filter at 780~nm is 
inserted to block possible fluorescence photons at $\lambda_{D1}=795$~nm.


\subsection{Intensity and decay of the redirected emission\label{secDensity}}

The ${\bf k}$-vector shift of dipole spin waves allows us to access both the subradiant states with strongly phase-mismatched wave vectors $|{\bf k}_s'|\neq \omega_{eg}/c$, and the redirected superradiant states with phase-matched wave vectors $|{\bf k}_s|= \omega_{eg}/c$. A study on the dynamics of subradiant states is left for a future publication~\cite{shortPaper2}. In this paper, we primarily focus on the dynamics of the directional superradiance. In particular, the redirected  superradiant emission from ${\bf k}_p$ to ${\bf k}_s$ in our setup naturally avoids the large probe excitation background, which commonly exists in previous studies of forward emission~\cite{superradianceRoof2016, Jun2016}, and thereby enables us to characterize the redirected superradiance with high accuracy. With the spin-wave excitation of timed-Dicke states prepared in a nearly ideal way (Sec.~\ref{secSimple}), we go beyond Ref.~\cite{coSub} and verify the $i_{{\bf k}_p}\propto N^2$ scaling as in Eq.~(\ref{equi2}). Furthermore, we observe possible deviation of the collective decay rate from $(1+\overline{{\rm OD}}/4)\Gamma$~\cite{clScullyPRL09,Kaiser2016, Sutherland2016,ZhuPRA2016}, which is likely related to a subtle superradiance reshaping effect~\cite{superradianceCottier18}.

\begin{figure}
\includegraphics[width=0.40\textwidth]{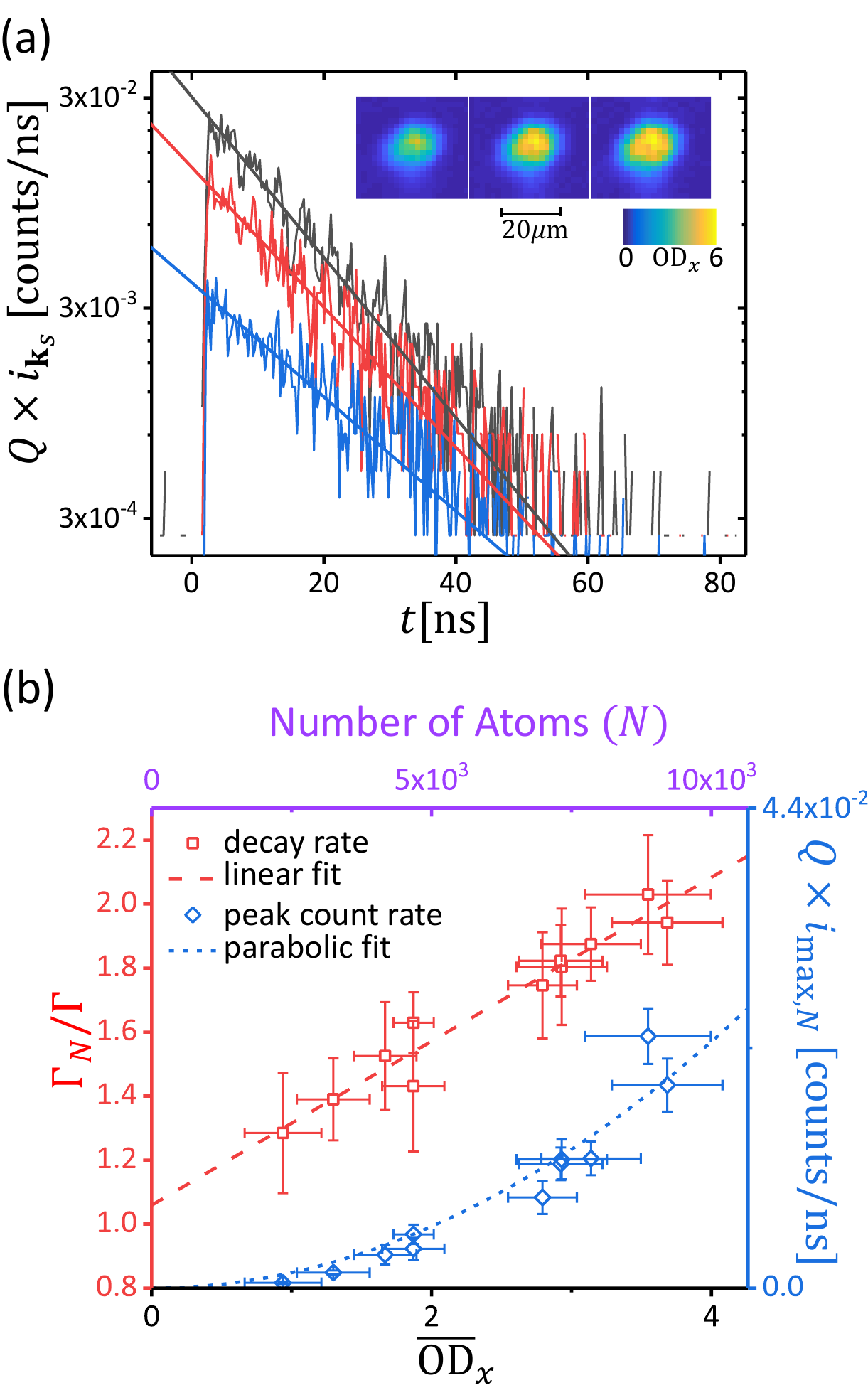}
\caption{ (a) Typical time-dependent redirected superradiant emission rate 
$i_s(t)$ multiplied by a detection quantum efficiency $Q\approx 0.15$, 
estimated over $N_{\rm exp}=13000$ experiments. The plot is on log scale. 
The curves are for samples with different atom numbers, with resonant 
optical depth images along ${\bf e}_x$ (insets) inferred from absorption 
images taken at the corresponding experimental conditions. The exponential 
fit gives the peak count rate $i_{{\rm max},N}$ and the collective decay 
rate $\Gamma_N$. (b) $i_{{\rm max},N}$ and $\Gamma_N$ are plotted versus 
estimated atom number $N$ and average optical depth $\overline{{\rm OD}}_x$. The 
error bars represent the statistical and fit uncertainties of the 
data.}\label{figDensity}
\end{figure}

We note Eq.~(\ref{equi2}) can describe what would be observed in the 
redirected emission in an ideal experiment, e.g., if the first pair of 
control pulses could be applied immediately ($\Delta t_1=0$) and perfectly 
after the probe pulse. To quantitatively describe the actual experiment, 
however, we must account for the fact that the spin wave already begins 
decaying in a superradiant fashion along the direction ${\bf k}_p$ during 
the delay time $\Delta t_1$, that the ensemble is non-spherical and has 
different optical depths ${\rm OD}_p$, ${\rm OD}_s$ along the directions ${\bf 
k}_{p,s}$, and that the control efficiency $f_d<1$ for the applied unitary 
operation of Eq.~({\ref{equUc}) is not perfect. The non-ideal $U_g$ control 
in presence of, e.g., $m_F$-dependent hyperfine phase shifts and spontaneous emission during the control, only partly converts the $S^{+}({\bf k}_p)$ into $S^{+}({\bf k}_s)$ 
excitation and further induces sub-wavelength density modulation in 
$\varrho({\bf r})$, thus we expect simultaneous and Bragg-scattering coupled 
superradiant emission into both the ${\bf E}_s({\bf k}_p)$ and ${\bf 
E}_s({\bf k}_s)$ modes. Practically for optical control with the focused 
laser beam as in this work,  we numerically find the ground-state atoms not 
shifted in momentum space are often associated with dynamical phase 
broadening, leading to suppressed $S^{+}({\bf k}_p)$ excitation and 
distorted sub-wavelength density fringes. A simple model for the superradiant photon emission rate into the redirected ${\bf E}_s({\bf k}_s)$ mode is 
\begin{equation}
i_s(t)\approx f_d \frac{\overline{{\rm OD}}_s}{\overline{{\rm OD}}_p} i_p(\Delta t_1) 
e^{-(1+(1-l) \overline{{\rm OD}}_s/4)\Gamma_{D2} (t-\Delta t_1)}.\label{equi4}
\end{equation}
Here, we have accounted for the different optical depths along the ${\bf k}_{s,p}$ directions, the finite control efficiency $f_d$, the superradiant decay that already occurs during the time $\Delta t_1$ along the original direction ${\bf k}_p$, and the fraction of atoms $l$ that are lost during the control process.

Experimentally, the spin-wave control for superradiance redirection with a ${\bf k}$ vector shift ${\bf k}_p \rightarrow {\bf k}_p - 2{\bf k}_c$ by a single pair of control pulses is first demonstrated  in Fig.~\ref{figDensity}(a), where the ${\bf E}_s({\bf k}_s)$ emission at a fixed delay $\Delta t_1=0.2$~ns is recorded. To study the collective effect of forward emission, we vary the atom number $N$ for atomic samples loaded into the same dipole trap with nearly identical spatial distribution. The time-dependent photon emission rate $i_s(t)$, obtained by normalizing the fluorescence counts with the number of runs $N_{\rm exp}$, counter time bin $\delta t$, and an overall detection quantum efficiency $Q\approx 
0.15$, nicely follows exponential decay curves for the accessed $N$ between $2\times 10^3$ and $9\times 10^3$ in this work. We extract both the peak emission rate $i_{{\rm max}, N}$ and collective decay rate $\Gamma_{N}$ with exponential fits, and study both quantities as a function of atom number $N$.

The cooperative nature of the collective emission is clearly demonstrated in 
Fig~\ref{figDensity}(b) with the $i_{{\rm max},N}\propto N^2$ scaling since 
according to Eqs.~(\ref{equi2}) and (\ref{equi4}) we have $i_{{\rm max}, 
N}\propto N\overline{{\rm OD}}_s$ but $\overline{{\rm OD}}_s\approx N\sigma_r/\sigma^2$ 
for our nearly spherical sample with size $\sigma$. We experimentally 
extrapolate the average optical depth $\overline{{\rm OD}}_s$ with ${\rm OD}_x(y,z)$ 
measurements along the $x$ direction [Fig.~\ref{figDensity}(a) insets, see 
Appendix~A for imaging details]. We have $\overline{{\rm OD}}_s= 
\xi\times \overline{{\rm OD}}_x$ with $\xi\approx 0.8$ to account for the ratio of 
optical depth integrated along the ${\bf k}_s$ and ${\bf e}_x$ directions 
respectively. By comparing the quadratic fit that gives $i_{{\rm 
max},s}\approx 4\times 10^{-4} N \Gamma_{D2} \overline{{\rm OD}}_x/4$ in 
Fig.~\ref{figDensity}(b) with Eqs~(\ref{equi2}) and (\ref{equi4}), we find a time-integrated probe Rabi frequency of  
$\theta_p \approx 2\times 10^{-2}$, taking our best understanding of the efficiency $f_d=0.7$ (see 
Sec.~\ref{secefficiency}). This value of $\theta_p$ is consistent with the expected 
excitation by the probe (with peak saturation parameter $s\sim 1$ and duration $\tau_p=$5~ns) in these 
measurements~\cite{foot:probenormalization}, considering the large 
uncertainty in the absolute intensity parameter estimations.

We now discuss the enhanced decay rate $\Gamma_N$ of the collective emission, which 
is approximated in Eqs.~(\ref{equi2}) and (\ref{equi4}) under the assumption of 
negligible angular-dependent emission dynamics (Appendix~B}). This 
approximate decay rate corresponds to that of the timed Dicke 
state~\cite{clScullyPRL09,Kaiser2016, Sutherland2016,ZhuPRA2016}. Similar to 
previous studies of forward 
superradiance~\cite{superradianceAraujo16,superradianceRoof2016}, 
we find $\Gamma_N\propto N$ for the redirected superradiance, as expected. Here, to make a precise comparison with the 
theoretical picture, we plot the same data in Fig.~\ref{figDensity}(b) versus the
{\it in situ} measured average optical depth $\overline{{\rm OD}}_x$. From 
Fig.~\ref{figDensity}(b), we have $\Gamma_N/\Gamma_{D2} \approx 1.1 +0.26 
~\overline{{\rm OD}}_x$. Using Eq.~(\ref{equi4}) again with 
$\xi=\overline{{\rm OD}}_s/\overline{{\rm OD}}_x\approx 0.8$ and the remaining fraction of atoms 
$(1-l)\approx 0.9$ in these measurements, as discussed in 
Sec.~\ref{secefficiency}, we obtain $\Gamma_N\approx (1.1+ \nu 
\times(1-l)\times\overline{{\rm OD}}_s)\Gamma_{D2}$ with $\nu=0.35\pm 0.1$, with 
no freely adjustable parameters but with an uncertainty limited by  the 
$\overline{{\rm OD}}_s$ estimation in this paper. The likely discrepancy between 
this result and the $\nu=0.25$, $\Gamma_N/\Gamma = 1+ \overline{{\rm OD}}/4$ 
prediction of the collective decay of the timed Dicke state~\cite{ZhuPRA2016, superradianceAraujo16,superradianceRoof2016} can be expected, since the measured collective emission $i_s(t)$ is integrated over 
the $\sigma$-limited solid angle $\sim 1/(k_p\sigma)^2$ beyond the ``exact'' 
${\bf k}_s={\bf k}_p-2{\bf k}_c$ phase-matching condition, while the small 
angle scattering of ${\bf E}_s({\bf k}_s)$ by the sample itself generally affects the collective emission dynamics~\cite{superradianceCottier18}, thereby violating our assumptions to reach Eq.~(\ref{equi2}).  


\subsection {Reversible shift of spin-wave ${\bf k}$ vector\label{expObs}}

\begin{figure}
    \includegraphics[width=0.45\textwidth]{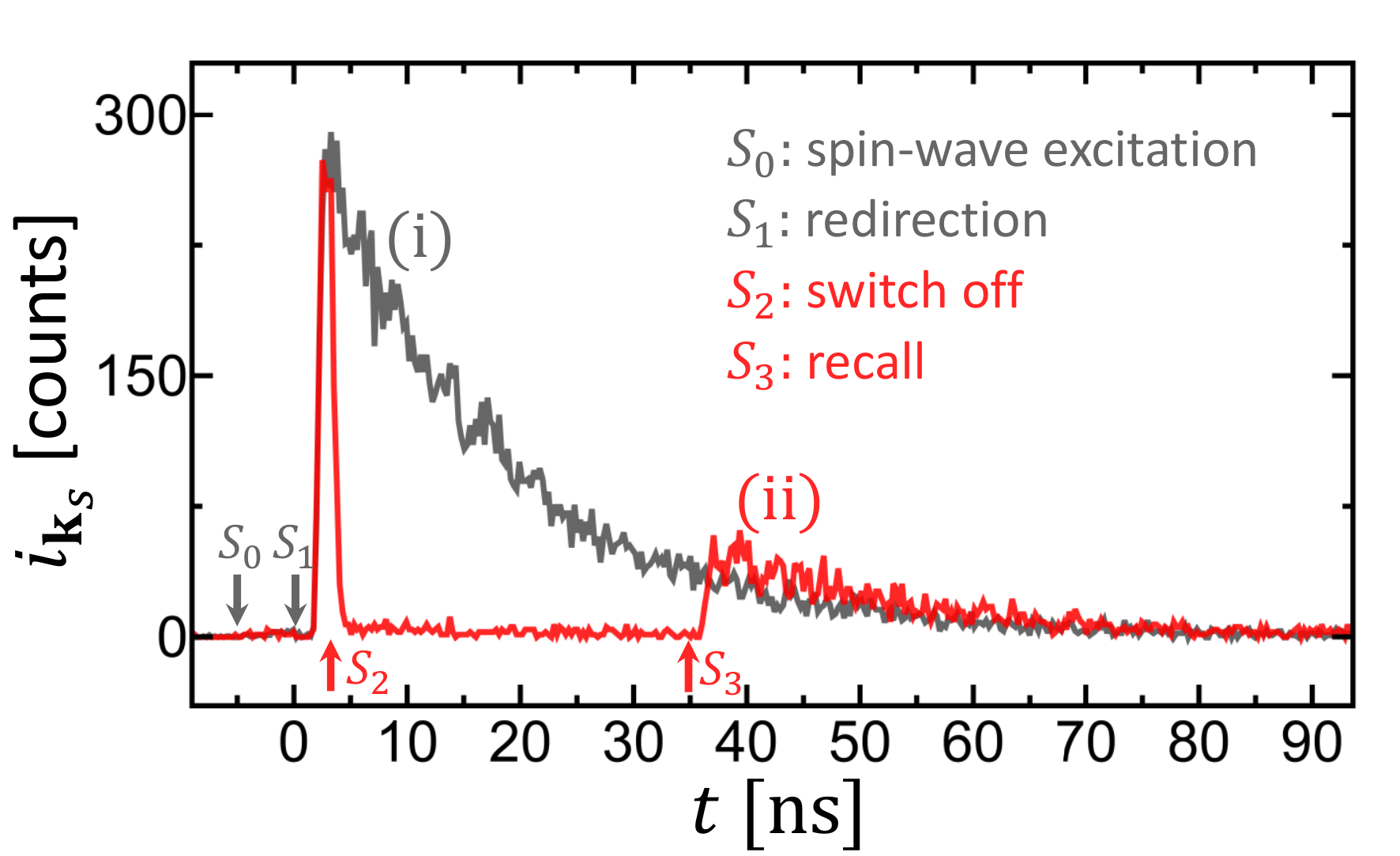}
    \caption{Experimentally measured photon emission counts $i_{{\bf k}_s}$ during spin-wave controls. In curve (ii), the collective spontaneous emission along ${\bf k}_s$ is switched on at $\Delta t_1 = 0.2$~ns, switched off at  $\Delta t_2 = 0.8$~ns, and recalled at $\Delta t_3 = 30.0$~ns. The pulses are optimized with $\tau_c=0.5$~ns, $\tau_d=1.2$~ns, $\Omega_0 \approx 2\pi\times 4$~GHz and $\delta_0 \approx 2\pi\times 4$~GHz. The spin-wave generation is implemented with a $\tau_p=5$~ns probe pulse. Here curve (i) shows the measured $i_{{\bf k}_s}$ under the same experimental conditions as those for the curve (ii), but for the redirection only and with the switch off and the recall operation removed.
        } \label{figsample}
\end{figure}

In the previous section, we have demonstrated the spin-wave control for the redirection operation, where a high-fidelity spin-wave vector shift ${\bf k}_p \rightarrow {\bf k}_p - 2 {\bf k}_c$ is implemented for the background-free detection of superradiance. We proceed further with our geometric control of the spin wave by applying a second pair of switch-off (${\bf k}_p - 2 {\bf k}_c \rightarrow {\bf k}_p - 4 {\bf k}_c$) control pulses along the ${\bf k}_c$ direction, followed by a third control for superradiance recall (${\bf k}_p - 4 {\bf k}_c \rightarrow {\bf k}_p - 2 {\bf k}_c$). The superradiant emission $i_{{\bf k}_s}$ in the ${\bf k}_s={\bf k}_p - 2 {\bf k}_c$ direction is recorded by the photon counter during the full redirection--switch off--recall sequence. We refer readers to Ref.~\cite{coSub} for discussions of the control sequence and the associated superradiance measurements. Here we provide an example signal $i_{{\bf k}_s}$ in Fig.~\ref{figsample}(ii), where the photon counts are initiated, shut off, and recalled at the expected instances when the three spin-wave control operations as in Fig.~\ref{figTimingsequence} are performed.


To understand the observations in Fig.~\ref{figsample}, we should note that after the switch-off control, the new wave vector ${\bf k}_s' = {\bf k}_p-4{\bf k}_c$ has 
is associated with wave number $|{\bf k}_s'|\approx 2.9~ \omega_{eg}/c$ that is strongly mismatched from radiation. For our dilute ensemble, this spin-wave excitation should decay in a superradiant-free fashion 
with approximately the single-atom decay rate $\sim\Gamma_{D2}$~\cite{scully2015}. At the same time, the emission into the 
same detection solid angle, without the collective enhancement, should decrease by a factor of $\sim N$~\cite{ZhuPRA2016}. Following the recall control, the spin wave is phase matched again and the superradiance is recalled at the desired time after a $\Delta t_3$ delay. It should be noted that the recalled superradiance decays almost on the same time scale of $\tau \approx 15$~ns as those for the directly redirected superradiance, as shown in the reference curve (i) under the otherwise identical experimental conditions. The reduced amplitude of the recalled emission (see Sec.~\ref{secSimpleDynamics}) reflects gradual decay of the mismatched ${S^+}({\bf k}_s')$ spin-wave order in the atomic gas (with an experimentally estimated lifetime $\sim 26$~ns in Ref.~\cite{coSub}), before its conversion back to the phase-matched ${S^+}({\bf k}_s)$ excitations. The fact that we can map the  phase-mismatched spin-wave order to light should have significant consequences when this technique is applied to arrays, where the subradiant dynamics has been predicted to be particularly rich~\cite{selectiveRadiancePRXchang17, Sutherland2016, syznanov2016, Perczel2017, Bettles2017}. 



\subsection{Optical acceleration}
As discussed in Sec.~\ref{secSimple}, the control pulse sequence to shift the spin waves also results in a spin-dependent kick, which optically accelerates the phase-patterned $|g\rangle$ states by the geometric force~\cite{geometricCheneau2008}. The momentum transfer along the control beam along ${\bf e}_z$ can 
be evaluated by integrating $\langle \hat F_z \rangle$ with the single-atom 
force operator $\hat F_z=-\frac{\hbar}{2} \partial_z \Omega_c 
|a\rangle\langle g|+{\rm H.c.}$, as the projected atomic state evolves on the 
$\{|g\rangle-|a\rangle\}$ Bloch sphere [Fig.~\ref{figTimingsequence}(b)]. For ideal 
population inversions, the integrated Berry curvature~\cite{Gritsev2012} 
gives the exact photon recoil momentum $\Delta P=2\hbar k_c$ with $\hbar$ 
the reduced Planck's constant.

\begin{figure} \includegraphics[width=0.35\textwidth]{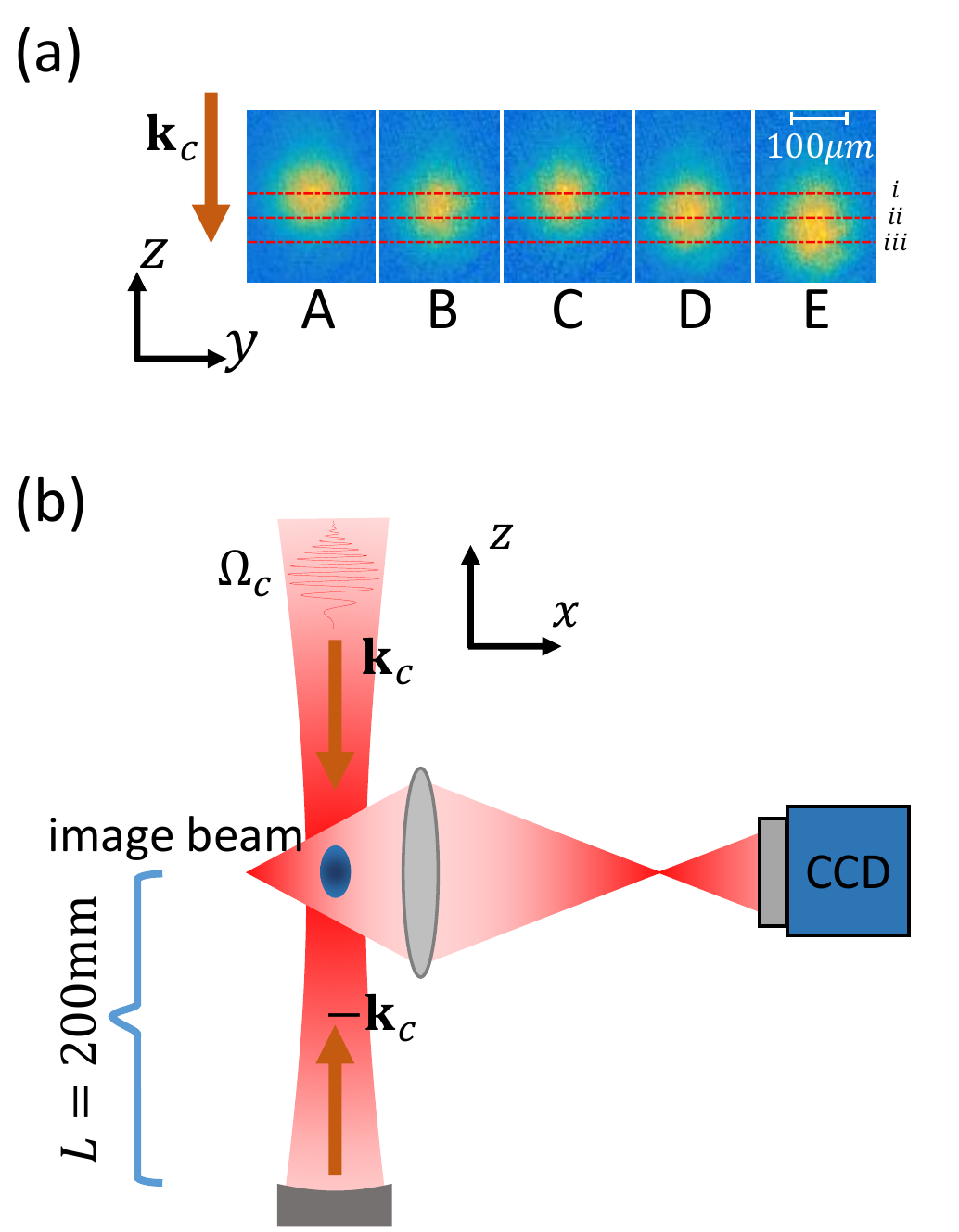} \caption{(a) 
Typical absorption images of $D1$ controlled atomic samples after a $T_{\rm tof}=400~\mu$s time-of-flight.  The three red dashed lines mark the expected positions for different momentum transfer: $\Delta P=0 
\hbar k_c$(i), $2\hbar k_c$(ii), and $4\hbar k_c$(iii).  For images A-D, each $D1$ control consists of a pair of chirp pulses with calibrated peak Rabi frequency $\Omega_0= 0$~GHz, $2\pi \times 0.8$~GHz, $2\pi \times 
1.9$~GHz, and $2\pi \times 2.7$~GHz and chirp parameter $\delta_0= 2\pi 
\times 0.1$~GHz, $2\pi \times 0.1$~GHz, $2\pi \times 0.1$~GHz, and $2\pi 
\times 3.4$~GHz, respectively. For image E, each $D1$ control consists of two 
pairs of chirped pulses with the same parameters as that for image D. The 
central positions of the atomic samples can be obtained with Gaussian fits. 
These parameter combinations are also marked in Fig.~\ref{figAcc}(a). (b) Imaging setup for the optical acceleration measurements. The 
setup is with aberration-free numerical aperture of NA$\approx$0.3. 
 } \label{figImg}
\end{figure}

\begin{figure*}
\includegraphics[width=1\textwidth]{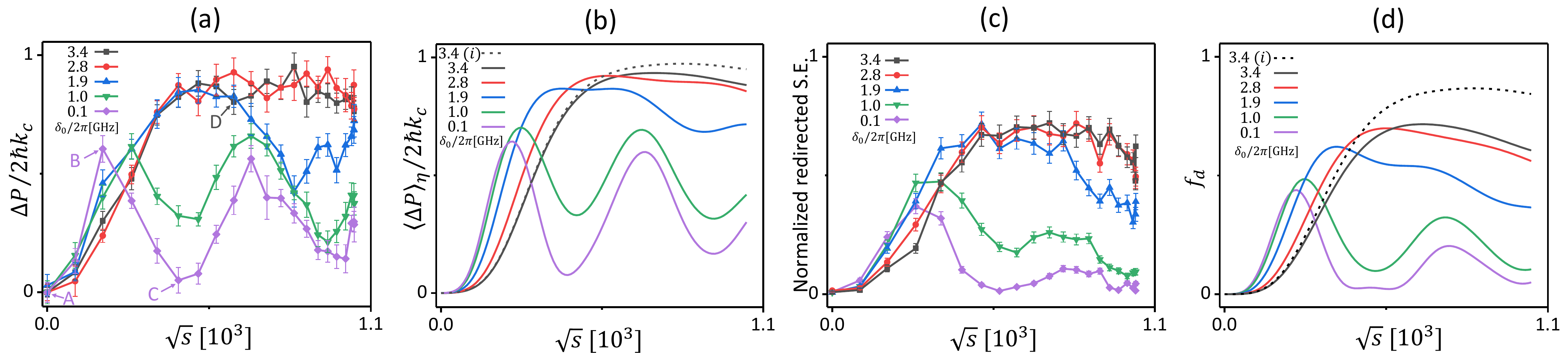}
\caption{Optical acceleration and efficiency for a ${\bf k}\rightarrow {\bf k}-2{\bf k}_c$ spin wave control with various pulse shaping parameters. Here $\tau_c=0.9$~ns,  $\tau_d=1.36$~ns. The mean momentum 
shift $\Delta P$ in (a) and normalized redirected superradiance (S.E.) in 
(c) are plotted versus peak laser intensity parameter $\sqrt{s}$ for control 
pulses with various chirping parameters $\delta_0$. (b) and (d) are 
simulated momentum transfers and dipole control efficiencies $f_d$ with 
additional laser pulse parameters optimally estimated. The simulation also 
provides acceleration and dipole control efficiency for an atom starting in the Zeeman state $m_F=0$ (dashed lines labeled with (i) in the legends). In (a), the A-D markers give parameters for absorption images presented in Fig.~\ref{figImg}.
}\label{figAcc}
\end{figure*}

Going beyond Ref.~\cite{coSub}, we experimentally measure the recoil momentum transfer $\Delta P$ associated with the $D1$ chirped 
pulse pair along ${\bf e}_z$ for the spin wave control, using a time-of-flight (TOF) absorption imaging method [Fig.~\ref{figImg}(b)]. Keeping in mind the Doppler effects due to the acceleration affect negligibly the 
nanosecond control dynamics, we repeat a ${\bf k}\rightarrow {\bf k}-2{\bf k}_c$ control pulse $N_{\rm rep}=5$ times to 
enhance the measurement sensitivity. The period $T_{\rm rep}=$ 440~ns $\gg 
1/\Gamma_{D1}$ is set to ensure independent interactions. We then measure the central position shift of the atomic sample after a $T_{\rm tof} = 400~ \mu$s TOF, using calibrated absorption images. For the absorption imaging, the atomic sample is illuminated by a $\tau_{\rm exp}=20~\mu$s imaging pulse resonant to 5$S_{1/2}$~$F=2 -  $5$P_{3/2}$~$F'=3$ along the ${\bf e}_x$ direction. The 2D transmission profile $T(y,z)=I/I_0$ is obtained by processing the imaging beam intensity $I,I_0$ on a CCD camera with and without the atomic sample, respectively. We then process $T(y,z)$ to obtain the optical depth profile ${\rm OD}(y,z)$ of the samples (also see Appendix A). Next, the central position shifts $\Delta z_0$ are obtained with 2D Gaussian fits for samples with and without the control pulses. Finally, a velocity change is estimated as $\Delta v=\Delta z_0/(T_{\rm tof}+\tau_{\rm exp}/2)$. Typical absorption images are presented in Fig.~\ref{figImg}(a) for samples under $D1$ control pulses with various  $\Omega_0$ and $\delta_0$ parameters as defined in Sec.~\ref{secResult}. In particular, we expect that the parameters used in image D in Fig.~\ref{figImg}(a) lie in an error-resilient region of parameter space, and reflect the nearly ideal momentum change of $\Delta P = 2\hbar k_c$. We shortly show that the measured momentum kicks agree well with numerical models.

With the knowledge of recoil velocity $v_r=\hbar k_c/m\approx 5.8~$mm/s, we obtain $\Delta P=\Delta v/N_{\rm rep} v_r\times \hbar k_c$ per $D1$ control. Typical $\Delta P$ measurements are presented in Fig.~\ref{figAcc}(a) versus intensity parameter $\sqrt{s}$, for shaped pulses with different chirping parameters 
$\delta_0$. For controls with nearly zero chirp ($\delta_0=2\pi\times 0.1, 
1.0$~GHz), $\Delta P$ displays a damped oscillation, which is due to optical 
Rabi oscillations with broadened periodicity associated with intensity 
inhomogeneity of the focused laser. The oscillation is suppressed at large 
$\delta_0$, with $\Delta P$ reaching $89(4)\%$ of the $\sim 2\hbar k_c$ 
limit at large $s$, suggesting a robustness to our coherent control process.  
Similar measurements are performed for the reversed ${\bf k}\rightarrow {\bf k}+2{\bf k}_c$ controls which results in opposite momentum shifts.

\subsection{Control efficiency: calibration and optimization 
\label{secefficiency}}

To quantify the imperfections in implementing the spin-wave control by geometric phase patterning 
[$U_g$ in Eq.~(\ref{equUc})], we need to properly model the dissipative 
dynamics of collective dipoles. For this purpose, we introduce the coherent 
dipole control efficiency, $f_d =\langle {\rm tr}(\rho_{\eta} S^+({\bf 
k}_s)S^-({\bf k}_s))\rangle_{\eta}/{\rm tr}(\rho_0 S^+({\bf k}_s)S^-({\bf 
k}_s))$, with $\rho_{\eta}$, $\rho_0$ the density matrix that describes the 
weakly $D2$ excited atomic sample subjected to the non-ideal $\tilde U_{g}$ 
and the ideal, instantaneous $U_g(\Delta {\bf k})$ control by Eq.~(\ref{equUc}), 
respectively. Here $\tilde U_{g}(\Delta {\bf k}; \Omega_0,\delta_0,\eta)$ due to the 
nanosecond shaped pulse control is parameterized by the peak Rabi frequency 
$\Omega_0$ and chirping parameter $\delta_0$, as well as factors $\eta_{1,2}$ 
as the normalized laser intensities for the forward and retro-reflected 
pulses locally seen by the atoms. The control efficiency $f_d$ is averaged over the Gaussian 
intensity distribution.

To optimize the ${\bf k}\rightarrow {\bf k}-2{\bf k}_c$ shift by the non-ideal $\tilde U_{g}(2 {\bf k}_c; \Omega_0,\delta_0,\eta)$ control, experimentally we simply scan the control pulse shaping parameters $\sqrt{s}\propto \Omega_0$ and $\delta_0$ to maximize the redirected superradiant emission that generate the time-dependent signal such as the curve (i) in Fig.~\ref{figsample}. The data in Fig.~\ref{figAcc}(c) are the corresponding total photon counts by integrating the time-dependent signals. By optimizing the total counts, we are able to locate the optimal pulse shaping parameters 
$\Omega_0=2\pi\times 2.7~$GHz and $\delta_0=2\pi\times 3.4$~GHz for the $\tau_c=0.9$~ns chirped-sine pulses in these experiments. 

The magnitude of the optimally redirected superradiant emission scales quadratically with total atom number $N$ and increases with both the probe excitation strength $|\Omega_p \tau_p|^2\ll 1$ and the control efficiency $f_d$ [see Eqs.~(\ref{equi2}) and~(\ref{equi4})]. However, without accurate knowledge of the experimental parameters associated with the spin-wave preparation and emission detection, it is difficult to precisely quantify $f_d$ with the photon counting readouts. Instead, we calibrate the spin-wave control efficiency $f_d$ with a numerical modeling strategy. In particular, we perform density-matrix calculations of a single atom interacting with the control pulses, including full hyperfine structure. Both the collective spin wave shifts and matter-wave acceleration can be evaluated from the single-atom results, if atom-atom interactions and re-scattering of the control fields is negligible, as we expect to be the case for the low atomic densities and high pulse bandwidths used in our work. For the calibration, we first adjust experimental parameters in numerical simulations so as to optimally 
match the simulated average momentum shift $\langle \Delta P\rangle_{\eta}$ 
in Fig.~\ref{figAcc}(b) with the absolute experimental measurements in Fig.~\ref{figAcc}(a). The corresponding $f_d$ under identical experimental conditions are then calculated as in Fig.~\ref{figAcc}(d). The fairly nice match between the superradiance measurements in Fig.~\ref{figAcc}(c) and Fig.~\ref{figAcc}(d) is achieved by uniformly normalizing the total counts in Fig.~\ref{figAcc}(c), with no additionally adjusted parameters. Near the optimal control regime, the simulation suggests we have reached a collective dipole control efficiency $f_d = 72\pm 4\%$, accompanied with the observed $f_a\equiv \Delta P/2\hbar k_c=89\pm 4\%$ acceleration efficiency. Constrained by the absolute acceleration measurements, we found this optimal $f_d$ estimation to be quite robust in numerical modeling when small pulse-shaping imperfections are introduced.


On the other hand, for a full redirection--switch off--recall sequence as in Fig.~\ref{figsample}, we can also estimate the efficiency of the ${\bf k}\rightarrow {\bf k}+2{\bf k}_c$ recall. We fit the amplitude of the recalled superradiant emission in the short $\Delta t_3$ limit, and compare that with the amplitude of the redirected emission with the same experimental sequence~\cite{coSub}. The ratio between the two fluorescence signal amplitudes defines an overall storage-recall efficiency for the controlled dipole spin wave intensity of $\sim 58\%$. By assuming equal efficiency for each of the two operations, the efficiency for a single ${\bf k} \rightarrow {\bf k} \pm 2 {\bf k}_c$ shift is thus at the 75(5)\% ($\sim \sqrt{58 \%}$ ) level, which are performed using the second type of delay line [Fig.~\ref{figTwodelayline}(b)] and reoptimized pulse-shaping parameters. The efficiency is also consistent with the prediction by numerical modelings, as discussed in Appendix~\ref{AppA}.

The optimal $f_d$ as in Fig.~\ref{figAcc}(c) is limited by $m_F$-dependent hyperfine phase shifts and 
$D1$+$D2$ spontaneous decays during the $\tau_d+\tau_c=2.26$~ns control. In 
particular, a $l\sim 10\%$ atom loss due to $D1$ spontaneous emission and 
$5P_{1/2}$ population trapping (particularly for $|m_F|=1$ states) is 
expected to reduce the number of atoms participating in the $D2$ collective 
emission. With atoms prepared in a single $m_F=0$ state, spontaneous 
emission limited dipole control efficiency of $f_d\approx 87\%$, accompanied with 
an acceleration efficiency $f_a\approx 97\%$ should to be reachable 
[Figs.~\ref{figAcc}(b) and \ref{figAcc}(d)] with the same control pulses.

\section{Discussions}
The error-resilient state-dependent phase patterning technique demonstrated 
in this paper is a general method to precisely control dipole spin waves in atomic gases and 
the associated highly directional collective spontaneous emission in the time 
domain~\cite{scully2015, foot:longlived, foot:superLattice, foot:otherField}. The control is itself a single-body technique, which can 
be accurately modeled for dilute atomic gases when the competing resonant dipole-dipole interactions between atoms can be ignored during the pulse duration. With the geometric phase inherited from the optical phases of the control laser beams, it is straightforward to design 
$\varphi_{\rm G}$  beyond the linear phase used in this paper and to manipulate the 
collective spin excitation in complex ways tailored by the control beam 
wavefronts. 

We note that the atomic motion associated with the control can limit the coherence time of the spin-wave order in free gases at finite temperature, and that the limiting effect can be suppressed with optical lattice confinements~\cite{coSub}. In the following, we discuss methods for perfecting the spin-wave control in dense gases, and then summarize possible prospects opened by this work.

\subsection{Toward perfect control with pulse shaping\label{secLimits}}

The optical dipole spin-wave control in this work is subjected to various imperfections at 
the single-body level. The pulse-shaping errors combined with laser 
intensity variations lead to imperfect population inversions and reduced 
operation fidelity. The imbalanced beam pair intensities lead to spatially 
dependent residual dynamical phase writing and distortion of the collective 
emission mode profiles. The hyperfine coupling of the electronically excited 
states may lead to inhomogeneous phase broadening as well as hyperfine Raman 
couplings, resulting in coherence and population losses as in this work. 
Finally, the spontaneous decays on both the $D1$ control and $D2$ probe channels 
limit the efficiency of the finite-duration pulse control. However, the 
imperfections of the control stemming from the single-atom effects are 
generally manageable with better quantum control 
techniques~\cite{StAChen2010,freegarde2014,NatCommDu2015,Chuang2016} 
well-developed in other fields, if they can be implemented in the optical 
domain with reliable pulse-shaping systems of sufficient precision, bandwidth, 
and output power. 

Beyond single-body effects, we emphasize that with the increased $\Omega_c$  strength and reduced $\tau_{c,d}$ time, it is generally possible to suppress interaction effects so as to maintain the precision enabled by 
the single-body simplicity, for precise spin-wave control in denser atomic gases.

As discussed in Sec.~\ref{secefficiency}, the pulse-shaping system used in this work already supports $f_d\sim 87\%$ efficiency 
if atomic $m_F$ states are better prepared, which is then limited by the  $D1$ and $D2$ spontaneous decay (single-atom limit) during the 
$\tau_c+\tau_d=2.26$~ns control time. Instead of imparting geometric phase to the ground-state atoms, in future work an $|e\rangle-|a\rangle$ transition with a longer $|a\rangle$ lifetime~\cite{multiphotonMOT} may be chosen to implement a $U_e(\varphi_{\rm G})$ for $|e\rangle$-state phase-patterning. The influence from the $D1$ decay can thus be eliminated, leading to $f_d\sim 95 \%$ limited by the suppressed $D2$ decay. With an additional fivefold reduction of $\tau_c$ to enable
$\tau_c+\tau_d$ to $\sim 400~$ps, aided by the well-developed advanced 
error-resilient techniques~\cite{StAChen2010,freegarde2014,NatCommDu2015,Chuang2016} , we expect $f_d$ reaching $99\%$ for high-fidelity dipole spin-wave control.

For the error-resilient shaped optical pulse control, ideally the fivefold reduction 
of control time from the $\tau_c=0.5-0.9$~ns pulses in this work needs to be supported by a fivefold increase of laser modulation bandwidth and a 25-fold increase of laser intensity.
Starting from the subnanosecond pulse-shaping technique in this paper detailed in 
Ref.~\cite{coSub}, the improvement may be achieved with a combined effort of stronger input, wider modulation and tighter laser focus. As a promising alternative, the control pulses may also be generated with mode-locked 
lasers~\cite{Freegarde1995,Immanuel1997,slowCampbell2014,campbell2019,Monroe2007,Yudi2020} with orders of magnitudes enhanced peak power and pulse bandwidth. Here we notice that for the same control operation, the required pulse peak power and energy scales with $1/\tau_c^2$ and  $1/\tau_c$, respectively. For controlling macroscopic samples as in this work, the scaling toward ultrafast pulses can become demanding enough to require sophisticated optical pulse amplifications, compromising the setup flexibility. In addition, the control strength $\Omega_c \gg 1/\tau_c$ and the modulation bandwidth are also upper-bounded too to 
minimize uncontrolled light shifts and multi-photon excitations.  Therefore, for the purpose of precisely and flexibly controlling dipole spin waves, it appears shaping picosecond pulses is more preferred than shaping ultrafast pulses ~\cite{slowCampbell2014,Yudi2020,shaperTutorial2010,Weiner2011Review} for generating nearly resonant pulses with a suitable duration and modulation bandwidth. 


\subsection{Summary and outlook}
In this paper, we experimentally demonstrate and systematically study a state-dependent geometric phase patterning technique for control of collective spontaneous emission by precisely shifting the ${\bf k}$ vector of dipole spin waves in the time domain. The method involves 
 precisely imparting geometric phases to electric dipoles in a large sample, using a focused laser beam with large intensity inhomogeneities. Similar error-resilient techniques have been widely applied in nuclear magnetic resonance~\cite{NMRAdiabatic2001,NatCommDu2015,
geometricAspect2012, Chuang2016}. Our work represents a step in exploring such error resilience toward optical control of dipole spin waves near the unitary limit, and for efficient far-field access to the rarely explored phase-mismatched optical spin-wave states.  During the characterization of our method, we also made intriguing observations related to fundamental properties of spin-wave excitations. These include a verification of the $i_N\propto N^2$ scaling law, a qualification of the enhancement relation $\Gamma_N/\Gamma=1 + \overline{{\rm OD}}/4$, and an observation of matter-wave acceleration accompanying the spin-wave control. We have provided a theoretical analysis of this spin wave and spontaneous 
emission control. Instead of working with free-space and randomly positioned atoms, our control technique can be readily applied to atomic arrays for efficient access to highly subradiant states. The technique may open the door to related applications envisaged in the field of quantum optics~\cite{scully2015,prl2016}, to help unlock non-trivial physics of long-lived and interacting dipole spin-waves in dense atomic gases~\cite{selectiveRadiancePRXchang17, DEC2019PRA, Sutherland2016,Guimond2019,Jennewein2016,oeJavanainen2017, Corman2017,syznanov2016,Perczel2017,Bettles2017, Molmer2019}, and to enable nonlinear quantum optics based on subradiance-assisted resonant dipole interaction~\cite{CarusottoRMP2013Rev,DarrickNP2014Rev,ChangsukRPP2017Rev}.

Finally, on the laser technology side, we hope this paper motivates additional developments of 
continuous and ultrafast pulse-shaping methods for optimal quantum control of optical electric dipoles.

\begin{acknowledgments}

We are grateful to Prof. Lei Zhou for both helpful discussions and for kind support and to Prof. J.~V.~Porto and Prof. Da-Wei Wang for helpful discussions and insightful comments on the paper. We thank Prof. Kai-Feng Zhao and Prof. Zheng-Hua An for help on developing pulse-shaping technology and for support from the Fudan Physics nanofabrication center. D.~E.~C. acknowledges support from the European Union's Horizon 2020 research and innovation program, under European Research Council Grant Agreement No. 639643 (FOQAL) and Grant Agreement No. 899275 (DAALI), MINECO Severo Ochoa Program No. CEX2019-000910-S, Generalitat de Catalunya through the CERCA program and QuantumCat (Reference No. 001-P-001644), Fundacio Privada Cellex, Fundacio Mir-Puig, Fundacion Ramon Areces Project CODEC, and Plan Nacional Grant ALIQS funded by Ministerio de Ciencia, Innovaci\'on y Universidades, Agencia Estatal de Investigaci\'on, and European Regional Development Fund. This research is mainly supported by National Key Research Program of China under Grants No. 2016YFA0302000 and No. 2017YFA0304204, by NSFC under Grant No. 11574053, and by Shanghai Scientific Research Program under Grant No. 15ZR1403200.

\end{acknowledgments}

\appendix
\renewcommand*{\thesection}{\Alph{section}}

\section{Experimental Details}

\subsection{Resonant ${\rm OD}$ and atom number measurements\label{secApImg}}
The absorption imaging setup as schematically illustrated in 
Fig.~\ref{figImg} not only helps us to quantify the optical acceleration 
effect with TOF technique, but also to directly measure the optical depth 
profile ${\rm OD}_x(y,z)$ and atom number $N$ as in Sec.~\ref{secDensity}. To 
investigate the $\Gamma_N/\Gamma=1+ \overline{{\rm OD}}/4$ relation, extra care 
was taken to extract the ${\rm OD}_x(y,z)$ images from the resonant absorption 
images. Here ${\rm OD}_x(y,z)$ to be measured should be the unpolarized atoms in 
the weak excitation limit, with {\it in situ} $\varrho({\bf r})$ distribution 
close to those in the quantum optics experiments and for both low 
$\overline{{\rm OD}}<1$ and quite high $\overline{{\rm OD}}\sim3.5$. To ensure 
consistent $\varrho({\bf r})$ distribution to be measured, a short exposure 
time of $20~\mu$s is chosen. To collect sufficient counts on the camera, we 
use imaging beams with quite high intensity in the range of $I_0=1\sim 
20$~mW/cm$^2$ and thus with a saturation parameter $s=0.3\sim7$ assuming 
$I_s = 3.05$~mW/cm$^2$~\cite{Steck2015} for $\pi$ transition of $5S_{1/2} 
F=2 - 5P_{3/2} F'=3$. We reduce the measurement uncertainty related to 
saturation effects following techniques similar to 
Refs.~\cite{ReinaudiOptLett2007,Chin2017}. In addition, to avoid measurement 
uncertainty related to low local counts for the highly absorbing samples, we 
calibrate the peak ${\rm OD}$ of the {\it in situ} samples with TOF images at reduced ${\rm OD}$. The processes are detailed as follows.

We start by repeated absorption imaging measurements for nearly identical 
TOF samples with 2D transmission profile $T(I)=I/I_0>75\%$, with incoming 
$I_0(y,z)$ and transmitted $I(y,z)$ intensities recorded on the camera. The 
optical depth profile in the weak excitation limit can be approximately as 
${\rm OD}_x(y,z)=-{\rm log} T(I)+(I_0-I)/I_s^{\rm 
eff}$~\cite{ReinaudiOptLett2007,Chin2017}. Here $I_s^{\rm eff}$ is an 
effective parameter for calibrating our saturation intensity measurements. 
By globally adjusting $I_s^{\rm eff}$ and thus the $(I_0-I)/I_s^{\rm eff}$ 
term, we obtain consistent ${\rm OD}_x(y,z)$ from all the measurements with 
$I_0=1\sim 20$~mW/cm$^2$ with minimal variations. Notice that the radiation 
pressure during the imaging process does not significantly vary the 
power-broadened atomic response.

The optimally adjusted $I^{\rm eff}_s$ serves to extract the ${\rm OD}_x(y,z)$ 
spatial profile for atomic sample immediately after their release from the 
dipole trap, as in Fig.~\ref{figDensity}(a) with approximately identical 
spatial profiles. In addition, under the consistent atomic sample 
preparation conditions, we also measure the optical depth profile 
${\rm OD}_x'(y,z)$ and total atom number after a $430~\mu$s TOF. The 
TOF greatly reduces the peak linear absorption for the highest 
${\rm OD}$ sample here from the expected $95\%\sim 99\%$ level down to $15\%\sim 
25\%$, leading to more accurate estimation of integrated ${\rm OD}$ that is served 
to calibrate the {\it in situ} $\overline{{\rm OD}}_x$ measurements. To account for 
optical pumping effects that tend to increase the $F=2-F'=3$ light-atom 
coupling strengths, a factor of $0.85$~\cite{Steck2015} is multiplied to the 
extracted ${\rm OD}_x(y,z)$.

We finally adjust ${\rm OD}_x$ due to the imaging laser frequency noise in this 
work by up to $30\%$, according to the measured linewidth broadening of the 
TOF sample absorption spectrum, and then obtain $\overline{{\rm OD}}_s$ using the 
sample aspect ratio estimated by the auxiliary imaging optics along ${\bf 
e}_z$. These last two steps introduce the largest 
uncertainties into our $\overline{{\rm OD}}_s$ estimation. It is worth noting that 
the laser noise correction tends to reduce the $\nu$ value in 
Sec.~\ref{secDensity}. We use $\sigma_r=1.59\times 10^{-9}$~cm$^2$ for 
linearly polarized probe on $5S_{1/2}, F=2$ levels to estimate 
$N=\frac{1}{\sigma_r}\int {\rm OD}_x(y,z)dy dz$.

\section{Theoretical model and numerical simulation\label{AppA}}
 
This Appendix provides a theoretical background associated with the experimental observations for both this paper and Ref.~\cite{coSub}. First, in Appendix~\ref{secOD}, we provide a minimal model to explain the collective spontaneous emission by the controlled spin waves. Next, in light of short control time $\tau_{c,d}$ with negligible atom-atom interaction effects, we set up a single-atom model to explain the control of spin waves supported by the nearly non-interacting atoms in Appendix B2-B4. Different from Ref.~\cite{coSub}, here we outline the method to model the interaction between the atomic sample with the focused laser beam by estimating the most likely experimental parameters, so as to clearly understand the physical limitations behind the inefficiency of our spin-wave control. Finally, in Appendix B5 we discuss the influence of controlled spin-wave dynamics subject to hyperfine interactions during the superradiant recall operation.

\subsection{Collective spontaneous emission from a dilute gas of two-level 
atoms\label{secOD}}

We consider the interaction between $N$ two-level atoms with a resonant 
electro-magnetic field at wavelength $\lambda_p$ and frequency 
$\omega_{eg}$. With transition matrix element ${\bf d}_{eg}=d_{eg}{\bf e}_d$, the 
absorption cross section is given by $\sigma_r=k_p \alpha_i$ with 
$\alpha_i=2 |{\bf d}_{e g}|^2/{\hbar \Gamma}$, $\Gamma$ being the linewidth of the 
$|e\rangle-|g\rangle$ transition. The atomic ensemble follows an average 
spatial density distribution $\varrho({\bf r})=\langle \sum_i \delta({\bf 
r}-{\bf r}_i) \rangle$ that is assumed to be nearly spherical and smooth, in 
particular, $\varrho({\bf r})$ does not vary substantially on length scales 
other than that close to its characteristic radius $\sigma \gg \lambda_p$. 
We further restrict our discussion to intermediate sample size with $\sigma\ll c 
\tau$, with $c$ the speed of light and $\tau$ the shortest time-scale of 
interest. The transmission of a plane-wave resonant probe beam at the exit 
of the atomic sample, in the ${\bf r}=\{{\bf r}_{\perp},r_p\}$ coordinate, 
follows the Beer-Lambert law with transmission $T({\bf 
r}_{\perp})=e^{-{\rm OD}({\bf r}_{\perp})}$. The 2D optical depth distribution is 
given by ${\rm OD}=N\varrho_c({\bf r}_{\perp})\sigma_r$, $\varrho_c({\bf 
r}_{\perp})=\frac{1}{N }\int \varrho({\bf r})d r_p$ being the normalized column density.

To describe both the collective dipole dynamics and its collective 
radiation, we regard the small atomic sample as system and free-space 
optical modes as reservoir. The electric-dipole interaction can be 
effectively described by the many-atom density matrix $\rho$, after the 
photon degrees of freedom are eliminated by the standard Wigner-Weisskopf 
procedure. Following the general approach~\cite{Gross1982, selectiveRadiancePRXchang17}, 
the density matrix $\rho$ obeys the master equation  $\dot 
\rho=\frac{1}{i\hbar}(H_{\rm eff}\rho-\rho H^{\dagger}_{\rm eff})$ $+{\it 
L_c}[\rho]$, where $L_c$ is the population recycling superoperator associated with random quantum jumps in the stochastic wave-function picture. 
Here we focus on the effective Hamiltonian $H_{\rm eff}$ that governs the 
deterministic evolution of states and observables. The non-Hermitian 
effective Hamiltonian can be expressed as
\begin{equation}
H_{\rm eff}=\sum_i{H^i_a} +\hat V_{\rm D D,eff}, \label{eqHeff}
\end{equation}
with single-atom Hamiltonian $H^i_a$ for atom at location ${\bf r}_i$, and 
effective dipole-dipole interaction operator $\hat V_{\rm D 
D,eff}=\sum_{i,j}\hat V_{\rm D D}^{i,j}$ that sums over the pairwise 
resonant dipole interaction
\begin{equation}
    \hat V^{i,j}_{\rm D D}=-\frac{k_p^2}{\varepsilon_0} {\bf d}^*_{e g} \cdot
{\bf G}({\bf r}_i-{\bf r}_j,\omega_{eg}) \cdot {\bf d}_{e g} \sigma^+_i \sigma^-_j. 
\label{eqRDI}
 \end{equation}
Here $\sigma_i^+=|e_i\rangle\langle g_i|$, 
$\sigma_i^-=(\sigma_i^+)^{\dagger}$ are the raising and lowering operators 
for the $i^{\rm th}$ atom and $\varepsilon_0$ is the vacuum permittivity. 
${\bf G}({\bf r},\omega_{eg})$ is the free-space Green's tensor of the 
electric field obeying
\begin{equation}
\nabla\times\nabla\times {\bf G}({\bf r},\omega_{eg})-\frac{\omega_{eg}^2}{c^2} 
{\bf G}({\bf r},\omega_{eg})=\delta^3({\bf r}) \mathds{1}.
\end{equation}
Intuitively, Eq.~(\ref{eqRDI}) allows for the exchange of excitations 
between atoms, which is mediated by photon emission and re-absorption, and 
whose amplitude thus naturally depends on ${\bf G}({\bf r},\omega_{eg})$ which 
describes how light propagates from one atomic position to another.

With the spin-model description of the atomic dipole degrees of freedom, the 
electric field operator, describing the light emitted by the atoms, can be 
written in terms of the atomic properties as:
\begin{equation}
    \hat{\mathbf{E}}_s({\bf r})=\frac{k_p^2}{\varepsilon_0} \sum_i^N {\bf G}({\bf r}-{\bf 
r}_i,\omega_{eg}) \cdot {\bf d}_{e g} \sigma^-_i.
\end{equation}

Instead of generally discussing evolution of  atomic states in the $N-$spin 
space governed by $H_{\rm eff}$, in the following we discuss the timed-Dicke 
state $|\psi_{\rm TD}({\bf k})\rangle=S^+({\bf k})|g_1,g_2,\cdots,g_N\rangle$ 
and observables composed of collective linear operators. The results can 
then be straightforwardly applied to weakly excited gases in the linear 
optics regime as in this experiment.

We first consider the field amplitude of the spontaneously emitted photons. 
The emitted single photon from a timed Dicke state has a spatial mode 
profile $\boldsymbol{\varepsilon}_{\bf k}({\bf r})=\langle g_1,g_2,...,g_N|\hat {\bf 
E}_s({\bf r})|\psi_{\rm TD}({\bf k})\rangle$, which is readily re-written 
after the $\{{\bf r}_i \}-$configuration average as

\begin{equation}
   \overline{\boldsymbol{\varepsilon}_{\bf k}}({\bf r})= \frac{k_p^2  
\sqrt{N}}{\varepsilon_0} \int {\bf G}({\bf r}-{\bf r}',\omega_{eg}) \cdot {\bf d}_{e g} e^{i {\bf 
k}\cdot {\bf r}'}\varrho({\bf r}')d^3 r' .\label{equEmode1}
\end{equation}

Writing the spatial coordinate as $r=\{r_\perp,r_{\bf k}\}$, one can first 
integrate Eq.~(\ref{equEmode1}) at a fixed perpendicular coordinate over 
$r_p$, to obtain the emitted field at the end of the sample as in 
Eq.~(\ref{equEs}) with $\delta k=|{\bf k}|-\omega_{e g}/c$. The approximate 
integration assumes slowly-varying amplitude along both ${\bf r}_{\perp}$ 
and $r_{\bf k}$ directions. For ${\bf k}={\bf k}_p$ with $\delta k=0$, we 
then integrate the corresponding intensity over all transverse positions ${\bf 
r}_{\perp}$, and  normalize the radiation power by the energy 
$\hbar\omega_{eg}$ of a single photon to obtain the the collective photon 
emission rate $i_{{\bf k}_p}^{(1)}=\frac{2\varepsilon_0 c}{\hbar \omega_{eg}}\int 
|\overline{\boldsymbol{\varepsilon}_p}({\bf r})|^2 d^2 {\bf r}_{\perp} \approx 
\overline{{\rm OD}}_p\Gamma/4 $. For weakly excited coherent spin wave excitation, 
this emission rate is multiplied by $N\theta_p^2$ as in Eq.~(\ref{equi2}).

We now discuss time dependence of collective spontaneous emission described 
by Eq.~(\ref{equi2}) in the main text. The topic is related to a collective Lamb shift in a dilute atomic gas, an important and quite subtle effect well 
studied in previous work~\cite{clScullyPRL09,Manassah12}. 
To apply the general theoretical predictions to this paper, we 
explore the spin model~\cite{selectiveRadiancePRXchang17} to revisit the 
decay part of the problem, for the quite dense and small samples here.

We consider free gas evolution with $H_{\rm eff}=\hat V_{\rm DD,eff}$ and 
time-dependent field amplitude $\boldsymbol{\varepsilon}_p({\bf r,t})=\langle 
g_1,g_2,...,g_N|\hat {\bf E}_s({\bf r},t)|\psi_{\rm TD}({\bf k}_p)\rangle$, 
for $|{\bf r}-{\bf r}_i|\gg \lambda_p$ and with $\hat {\bf E}_s({\bf r},t)$ 
evolving according to Heisenberg-Langevin equation $\dot{\hat{ \mathbf{E}}}_s=\frac{1}{i\hbar}(\hat{\bf{E}}_s\hat{V}_{\rm{DD,eff}}-\hat{V}_{\rm{DD,eff}}^{\dagger} \hat{\bf{E}}_s)+\hat{f}$. With the Langevin force $\hat 
f$ being averaged to zero, for $|\psi\rangle=|\psi_{\rm TD}({\bf 
k}_p)\rangle$ we have
\begin{equation}
    \dot{\boldsymbol{\varepsilon}}_p({\bf r,t})=-i\langle g_1,g_2,...,g_N|\hat {\bf 
E}_s({\bf r},t))\hat V_{\rm DD,eff}|\psi_{\rm TD}({\bf 
k}_p)\rangle.\label{Eqt1}
\end{equation}

To evaluate Eq.~(\ref{Eqt1}), we insert the orthogonal timed-Dicke basis 
$\{|\psi_{\rm TD}({\bf k}_p)\rangle,|\psi_{1}({\bf 
k}_p)\rangle,...,|\psi_{N-1}({\bf k}_p)\rangle \}$ as in 
Ref.~\cite{clScullyPRL09} into the equation. Here $|\psi_{n}({\bf 
k}_p)\rangle=S_n^+({\bf k}_p)|g_1,...,g_N\rangle$ are single-excitation 
collective states with $S_n^+({\bf k}_p)=\sum_i 
c_{n,i}\sigma_i^+,n=1,...,N-1$ and with $c_{n,i}$ properly chosen to ensure 
the basis orthogonality~\cite{clScullyPRL09}. We further define the 
far-field emission amplitudes associated with the $N-1$ $|\psi_{n}({\bf 
k}_p)\rangle$ states as $\boldsymbol{\varepsilon}_{n}({\bf r},t)=\langle 
g_1,g_2,...,g_N|\hat {\bf E}_s({\bf r},t))|\psi_{n}({\bf k}_{p}\rangle$. We 
have,
\begin{equation}
\begin{array}{l}
    \dot{\boldsymbol{\varepsilon}}_p({\bf r},t)= -i  V_{\rm DD}({\bf k}_{p},{\bf k}_{p}) 
\boldsymbol{\varepsilon}_{p}({\bf r},t)\\
~~~~~~~~~~~~-i \sum_{n}  V_{\rm DD}(n,{\bf k}_{p}) \boldsymbol{\varepsilon}_{n}({\bf 
r},t),
\end{array}\label{Eqt2}
\end{equation}
with $V_{\rm DD}({\bf k}_{p},{\bf k}_{p})=\langle\psi_{\rm TD}({\bf k}_{p})| 
\hat V_{\rm DD,eff}|\psi_{\rm TD}({\bf k}_{p})\rangle$ and similarly $V_{\rm 
DD}(n,{\bf k}_{p})=\langle\psi_n({\bf k}_p)| \hat V_{\rm DD,eff}|\psi_{\rm 
TD}({\bf k}_{p})\rangle$. The second line of Eq.~(\ref{Eqt2}) includes random 
and collective couplings between the ${\bf k}_p$ superradiant excitation and 
other super- and sub-radiant modes~\cite{subradianceGuerin2016}, a fact 
associated with $|\psi_{\rm TD}({\bf k}_p)\rangle$ not being the eigenstate 
of $\hat V_{\rm DD,eff}$~\cite{clScullyPRL09,Manassah12, subRadianceKaiser2012}.

The $V_{\rm DD}({\bf k}_{p},{\bf k}_{p})\propto \sum_{i,j}{\bf d}^*_{e g} \cdot
{\bf G}({\bf r}_i-{\bf r}_j,\omega_{eg}) \cdot {\bf d}_{e g}e^{i{\bf k}_p \cdot ({\bf r}_i-{\bf r}_j)}$ factor in the 
first line of Eq.~(\ref{Eqt2}) is carefully evaluated as follows: For $i=j$, we have divergent ${\bf d}^*_{e g} \cdot {\bf G}(0^+,\omega_{eg}) \cdot {\bf d}_{e g}$ whose real part accounts for the single-atom Lamb shift and is absorbed into a redefinition of $\omega_{eg}$, 
with imaginary part equal to $\Gamma/2$ for isolated two-level atoms. The 
$i\neq j$ part is evaluated after the $\{{\bf r}_i\}-$configuration average 
as $V'= N \frac{k_p^2 |{\bf d}_{eg}|^2 }{\varepsilon_0} \int\int {\bf d}^*_{e g} \cdot {\bf G}({\bf r}-{\bf r}') \cdot {\bf d}_{e g} \varrho({\bf r})\varrho({\bf r}') e^{i{\bf k}_p \cdot ({\bf 
r}-{\bf r}')} d^3 {\bf r}d^3 {\bf r}'$. Following the same integration trick 
to arrive at Eq.~(\ref{equEs}), we rewrite this integration into the form of 
$V' \propto \int \overline{\boldsymbol{\varepsilon}_p}({\bf r}) \varrho({\bf r})$ to have
\begin{equation}
\begin{array}{l}
    V'\approx \frac{N \sigma_r \Gamma }{4 i}\int d^2 r_{\perp} \int d r_p 
\int^{r_p} d r_p' \varrho({\bf r}_{\perp},r_p') \varrho({\bf 
r}_{\perp},r_p),\\
~~~=-\frac{i}{8}\overline{{\rm OD}}_p\Gamma
    \end{array}\label{equVDDprime}
\end{equation}
with the normalized column density $\varrho_c({\bf r}_{\perp})=\int 
\varrho({\bf r}_{\perp},r_p) d r_p$ and with $\overline{{\rm OD}}_p=N \sigma_r\int 
\varrho_c({\bf r}_{\perp})^2 d^2 r_{\perp}$, as in the main text. We finally 
have

\begin{equation}
    \overline{V_{\rm DD}}({\bf k}_{p},{\bf k}_{p})\approx -\frac{i}{2} 
(1+\frac{\overline{{\rm OD}}_p}{4})\Gamma. \label{equVkk}
\end{equation}

To obtain the simple expression of $V'$ in Eq.~(\ref{equVDDprime}) and 
$\overline{V_{\rm DD}}({\bf k}_{p},{\bf k}_{p})$ in Eq.~(\ref{equVkk}), the 
SVE and Raman-Nath approximations are applied to evaluate 
$\overline{\boldsymbol{\varepsilon}_p}$  inside the sample. The approximations lead to 
field errors of order $\lambda_p/\sigma$ or higher. The corrections of these 
errors are associated with density-dependent corrections to 
Eq.~(\ref{equVkk}), including the collective Lamb 
shifts~\cite{clScullyPRL09}.

We come back to Eq.~(\ref{Eqt2}). For the smooth density distribution at 
moderate densities under consideration, the inter-mode couplings $V_{\rm 
DD}(n,{\bf k}_p)$ are generally expected to be quite weak and $\{ {\bf 
r}_i\}-$specific.
For the $\{ {\bf r}_i\}-$averaged fields, at an observation location ${\bf 
r}_o$ in the far field along the ${\bf k}_p$ direction, the couplings can be 
completely ignored initially, since with ${\bf G}({\bf r}_o-{\bf 
r}_i,\omega_{eg})\propto \frac{e^{i (k_p r_o-{\bf k}_p\cdot {\bf 
r}_i)}}{4\pi r_o}$ we have $\overline{\boldsymbol{\varepsilon}_p}({\bf r}_o,0)\propto \int\varrho({\bf r}) d^3 {\bf 
r}$ while $\overline{\boldsymbol{\varepsilon}_n}({\bf r}_o,0)=0$. We consider 
$\boldsymbol{\varepsilon}_p=\overline{\boldsymbol{\varepsilon}_p}+\delta\boldsymbol{\varepsilon}_p$, 
$\boldsymbol{\varepsilon}_n=\overline{\boldsymbol{\varepsilon}_n}+\delta\boldsymbol{\varepsilon}_n$,  $V_{\rm 
DD}=\overline{V_{\rm DD}}+\delta V_{\rm DD}$, and apply the $\{{\bf 
r}_i\}$-configuration average to  Eq.~(\ref{Eqt2}). By ignoring the 
$\overline{\boldsymbol{\varepsilon}_n}$ terms, we obtain the {\em initial decay} of 
$\boldsymbol{\varepsilon}_p({\bf r}_o,t)$ as
\begin{equation}
     \dot{\overline{ \boldsymbol{\varepsilon}_p}}({\bf r}_o,t)\approx -i 
\overline{V_{\rm DD}}({\bf k}_{p},{\bf k}_{p}) 
\overline{\boldsymbol{\varepsilon}_{p}}({\bf r}_o,t)+O(\langle \delta V \delta 
\boldsymbol{\varepsilon} \rangle).\label{Eqt2b}
\end{equation}

Equations~(\ref{equVkk}) and (\ref{Eqt2b}) suggest superradiant decay of 
directional spontaneous emission power at the $\Gamma_N = 
(1+\overline{{\rm OD}}_p/4)\Gamma$ rate on the exact forward (${\bf k}_p$) 
direction, for atomic samples at moderate densities ($N<k_p^3 \sigma^3$). 
Apart from predicting the decay rate of the far-field emission, it is worth 
pointing out that the $\Gamma_N = 2 {\rm Im} \langle \psi_{\rm TD}({\bf 
k}_p)| H_{\rm eff}|\psi_{\rm TD}({\bf k}_p)\rangle$ associated with 
Eq.~(\ref{equVkk}) is also applicable to the decay of $|\psi_{\rm TD}({\bf 
k}_p)\rangle$ population in the Schr{\"o}dinger 
picture~\cite{clScullyPRL09, Manassah12, subRadianceKaiser2012, ZhuPRA2016,Sutherland2016B, superradianceAraujo16,superradianceRoof2016}, and by energy conservation the initial rate of photon emission into $4\pi$. In this paper, we further 
approximately identify this decay rate with that for the observable 
$i_{{\bf k}_p}(t)\propto \int d^2 r_{\perp} |\overline{\boldsymbol{\varepsilon}_p}({\bf 
r}_{\perp},t)|^2$, leading to Eq.~(\ref{equi2}) in the main text for the 
collective emission. The same conclusion can be reached if one simply assumes the spatial profile $\boldsymbol{\varepsilon}_p({\bf r},t)$ would not change significantly during the 
emission, so as to ignore the  $V_{\rm DD}(n,{\bf k}_p)$ couplings. However, 
it is important to note that for $\boldsymbol{\varepsilon}_n$ in Eq.~(\ref{Eqt2}) 
associated with collective emission near the forward directions (close to 
${\bf k}_p$), the $V_{\rm DD}(n,{\bf k}_p)$ couplings can also be 
collective, and may strongly affect $\boldsymbol{\varepsilon}_p({\bf r},t)$ dynamics at 
${\bf r}$ along similar directions. Such couplings are just small angle 
diffractions by the averaged sample profile that generally lead to reshaped 
emission wavefronts $\boldsymbol{\varepsilon}_p({\bf r},t)$ over time~\cite{superradianceCottier18}, and, as a 
consequence, deviation of $i_{{\bf k}_p}(t)$ decay rate from that for the $|\psi_{\rm 
TD}({\bf k}_p)\rangle$ population. The last term in Eq.~(\ref{Eqt2b}) is 
associated with granularity of the atomic distribution, and we also expect 
that such granularity cannot be ignored for very high densities, or for 
systems with broken symmetry such as in a lattice.

In future work~\cite{shortPaper2}, it would be interesting to better understand the effect of 
discreteness on collective interactions, and in addition to investigate 
further the corrections due to the intermode coupling in Eq.~(\ref{Eqt2}) 
and the possible deviation from the dynamics of Eq.~(\ref{equi2}).

We remark that in all discussions in this paper, the replacement $|{\bf 
d}_{e g}|^2=\hbar \Gamma \alpha_i/2$ is general and applicable to atoms with 
level degeneracy. Thus we expect the conclusions for Eqs.~(\ref{equEs})-(\ref{equi4}) in 
the main text applicable to the $D2$ line of $^{87}$Rb atom in this paper.

\subsection{Simulation of spin-wave dynamics supported by
non-interacting atoms \label{SecExpModel}}

We now ignore the atom-atom interaction in Eq.~(\ref{eqHeff}), and write down the effective non-Hermitian Hamiltonian for the interaction-free model of $N$ three-level atoms as

\begin{equation}
  H'_{\rm eff}=\sum_{i=1}^N 
\big(H^i_a+H^i_e-i\hbar\frac{\Gamma_{D2}}{2}|e_i\rangle\langle 
e_i|-i\hbar\frac{\Gamma_{D1}}{2}|a_i\rangle\langle a_i| \big),\label{eqHeff2}
\end{equation}
where 
\begin{equation}
  H_a^i=-\hbar\Delta|a_i\rangle\langle a_i|+ \frac{\hbar}{2}\big(\eta({\bf 
  r}_i)\Omega_{c}(t) e^{-i\varphi_{c}({\bf r}_i, t)}\sigma^+_{c,i}+{\rm H.c.}\big)
   \label{Ha1}
\end{equation}
and 
\begin{equation}
  H_e^i=-\hbar\Delta_e |e_i\rangle\langle e_i|+ \frac{\hbar}{2}\big(\Omega_{p}(t) e^{-i\varphi_{p}({\bf r}_i)} \sigma^+_{i} +{\rm H.c.}\big).
   \label{He1}
\end{equation}
Here, ${\bf r}_i$ is the spatial position of the $i{\rm th}$ atom and we have $\sigma^+_{c,i}=|a_i\rangle\langle g_i|$ and $\sigma^+_{i}=|e_i\rangle\langle g_i|$. The control Rabi frequency is $\eta({\bf r}_i) \Omega_{c}(t)=|{\bf E}_c({\bf r}_i,t)\cdot {\bf d}_{a g}|/\hbar$, with a Gaussian beam intensity profile. $\Omega_c(t)$ is the spatially peak value of the control Rabi frequency and 
$\eta({\bf r}_i)\leq1$ is a position-dependent factor. $\Omega_{p}$ is the Rabi frequency of the probe pulse with $ \int_{-\tau_p}^{0} \Omega_p d t \ll 1$ as defined before. $\varphi_{p}({\bf r}_i) = -{\bf k}_p \cdot {\bf r}_i$ is the optical phase of the probe pulse. $\Omega_c(t)$ and $\dot{\varphi}_c({\bf r}_i,t) = \delta_c({\bf r}_i,t) $ are depicted in Fig.~\ref{figTimingsequence}(a) in the main text for robust state-dependent phase-patterning. By changing the basis into ${\bf k}$ space, we can rewrite the Hamiltonian in Eq.~(\ref{eqHeff2}) as $H_{\rm eff}=\sum H^{(s)}_{\rm eff}$ with
\begin{equation}
  \begin{array}{l}
H^{(s)}_{\rm eff}=H_p+H_{c1}+H_{c2},\\
\\
H_p=\hbar\sum_{g,{\bf k}}\Delta_g |g,{\bf k}\rangle\langle g,{\bf k}|\\
~~~~~~~+\hbar\sum_{e,{\bf k}} (-\Delta_e-i\Gamma_{D2}/2)|e,{\bf k}\rangle\langle 
e,{\bf k}|\\
~~~~~~~+\hbar\sum_{a,{\bf k}}(-\Delta_a-i\Gamma_{D1}/2)|a,{\bf k}\rangle\langle 
a,{\bf k}|\\
~~~~~~~+\hbar\sum_{g,e,{\bf k}}\big(\frac{1}{2}\Omega_{p}(t+\tau_p) c^y_{e g} 
|e,{\bf k}+{\bf k}_p\rangle \langle g,{\bf k}|+ {\rm H.c.} \big),\\
\\
H_{c1}=-\hbar\sum_{a,{\bf k}}\delta_c(t-t_1)|a,{\bf k}\rangle\langle a,{\bf k}|\\
~~~~~~~~+\hbar\sum_{g,a,{\bf k}}\big(\frac{1}{2}\eta_1 \Omega_{c1}(t-t_1) c^x_{a 
g} |a,{\bf k}+{\bf k}_c\rangle \langle g,{\bf k}|+ {\rm H.c.} \big),\\
\\
H_{c2}=-\hbar\sum_{a,{\bf k}}\delta_c(t-t_2)|a,{\bf k}\rangle\langle a,{\bf k}|\\
~~~~~~~~+\hbar\sum_{g,a,{\bf k}}\big(\frac{1}{2}\eta_2 \Omega_{c2}(t-t_2) c^x_{a 
g} |a,{\bf k}-{\bf k}_c\rangle \langle g,{\bf k}|+ {\rm H.c.} \big),
\end{array}\label{equHeff3}
\end{equation} 
where $|g,{\bf k}\rangle=\frac{1}{\sqrt{N}}\sum_i e^{i{\bf 
k}\cdot {\bf r}_i}|g_i\rangle$ and similarly for $|e,{\bf k}\rangle$ and 
$|a,{\bf k}\rangle$. Here we have included all the $D1$ and $D2$ hyperfine levels and use $\{g,e,a\}$ as indices to label the $\{5S_{1/2},5P_{3/2},5P_{1/2}\}$ hyperfine levels, respectively. The $c^x_{ag}$,$c^y_{eg}$ are coupling coefficients for ${\bf e}_x-$ and ${\bf e}_y-$ polarized $D1$ and $D2$ pulses, derived from the Clebsch-Gorden coefficients. Here $\eta_{1,2}$ are factors to account for the laser intensity inhomogeneities. Following the convention as in Fig.~\ref{figTimingsequence}(a), the probe excitation is between $-\tau_p<t<0$, which is followed by the two $D1$ control pulses ($H_{c1}$ and $H_{c2}$) starting at $t_1 = \Delta t_1$ and $t_2 = \Delta t_1 + \tau_d$, respectively. We finally end up with the master equation for the single-atom density matrix $\rho^{(s)}$ as
\begin{equation}
\begin{array}{l}
\dot{\rho}^{(s)}(t)=\frac{1}{i} (H^{(s)}_{\rm eff} \rho^{(s)} -\rho^{(s)} 
H_{\rm eff}^{(s)\dagger} )\\
~~~~~~~~~~~~~~~~~+\sum_{j} (\hat{C}^j_{D1} \rho^{(s)} 
\hat{C}_{D1}^{j\dagger}+\hat{C}^j_{D2} \rho^{(s)} \hat{C}_{D2}^{j\dagger}).
\end{array}\label{equiMasterS}
\end{equation}
The collapse operators are simply defined as 
\begin{equation}
  \begin{array}{l}
      \hat C^j_{D1}=\sum_{a,g,{\bf k}} \sqrt{\Gamma_{D1}}c^{j}_{a g} 
  |g,{\bf k}\rangle \langle a,{\bf k}|,\\
      \hat C^j_{D2}=\sum_{e,g,{\bf k}} \sqrt{\Gamma_{D2}}c^{j}_{e g} 
  |g,{\bf k}\rangle \langle e,{\bf k}+{\bf k}_p|,
  \end{array}\label{equiC}
  \end{equation}
with $j$ running through $x$, $y$, and $z$ polarizations. This effective evolution of the density matrix within the momentum lattice basis is summarized in Fig. 7.

With $\rho^{(s)}(t)$ it is straightforward to calculate the interaction-free 
evolution of the many-atom density matrix $\rho(t)=(\rho^{(s)}(t))^{\otimes N}$ 
and to evaluate collective observables $\langle \hat O \rangle={\rm 
tr}(\rho(t) \hat O)$~\cite{coSub}. By solving the master equation with the initial condition $\rho^{(s)} = \frac{1}{5}\sum_{g=1}^5 |g,{\bf k}\rangle\langle g,{\bf k}|$ (with $|g\rangle$ running through the $|F=2,m_F\rangle$ Zeeman sublevels), we further calculate the dipole coherence ${\bf d}({\bf k}_s) = {\rm tr}\big(\rho^{(s)} (t){\bf d}^{-}({\bf k}_s)\big)$, and similarly for ${\bf d}({\bf k}_p)$ and ${\bf d}({\bf k}_s')$. Here we define the operators ${\bf d}^{-}({\bf k}_s)={\bf e}_y \sum_{g,e}c^{y}_{e g}|g,{\bf k}+2 {\bf k}_c\rangle\langle e,{\bf k}+{\bf k}_p|$, ${\bf d}^{-}({\bf k}_p)={\bf e}_y \sum_{g,e}c^{y}_{e g}|g,{\bf k}\rangle\langle e,{\bf k}+{\bf k}_p|$ and ${\bf d}^{-}({\bf k}_s')={\bf e}_y \sum_{g,e}c^{y}_{e g}|g,{\bf k}+4 {\bf k}_c\rangle\langle e,{\bf k}+{\bf k}_p|$. The superradiant signal $i_{{\bf k}_s}$ in the main text is related to the expectation value $\langle S^{+}({\bf k}_s)S^{-}({\bf k}_s) \rangle$. In the large $N$ limit, we approximately have $\langle S^{+}({\bf k}_s)S^{-}({\bf k}_s) \rangle \approx |\langle S^{-}({\bf k}_s)\rangle|^2$, where $\langle S^{-}({\bf k}_s)\rangle$ is proportional to the dipole coherence $\langle {\bf d}^{-}({\bf k}_s) \rangle$ as $\langle S^{-}({\bf k}_s) \rangle\propto N\langle {\bf d}^{-}({\bf k}_s)\rangle$. Thus, by calculating the dipole coherence, the simulation can reproduce the $D2$ collective emission dynamics with $D1$ control for non-interacting atoms. With experimental imperfections encoded in parameters like $\eta_{1,2}$ in Eq.~(\ref{equHeff3}), we refer to the numerically evaluated single-atom density matrix according to Eq.~(\ref{equiMasterS}) as  $\rho^{(s)}_{\eta}(t)$. For comparison, the perfect geometric phase patterning is implemented by replacing the evolution by $H_{c1}+H_{c2}$ in Eq.~(\ref{equHeff3}) with instantaneous $U_g(-2{\bf k}_c)=1-\sum_g |g, {\bf k}\rangle \langle g,{\bf k}|+\sum_g |g,{\bf k}+2 {\bf k}_c\rangle \langle g,{\bf k}|$, leading to a perfectly controlled density matrix $\rho_0^{(s)}(t)$ for $t>0$. We then further define $i_{s,\eta}(t) = |{\rm tr}\big(\rho_{\eta}^{(s)}(t){\bf d}({\bf k}_s)\big)|^2$ and similarly $i_{s,0}(t) =  |{\rm tr}\big(\rho_0^{(s)}(t){\bf d}({\bf k}_s)\big)|^2$ for redirected superradiance under perfect control.

\begin{figure} \includegraphics[width=0.43\textwidth]{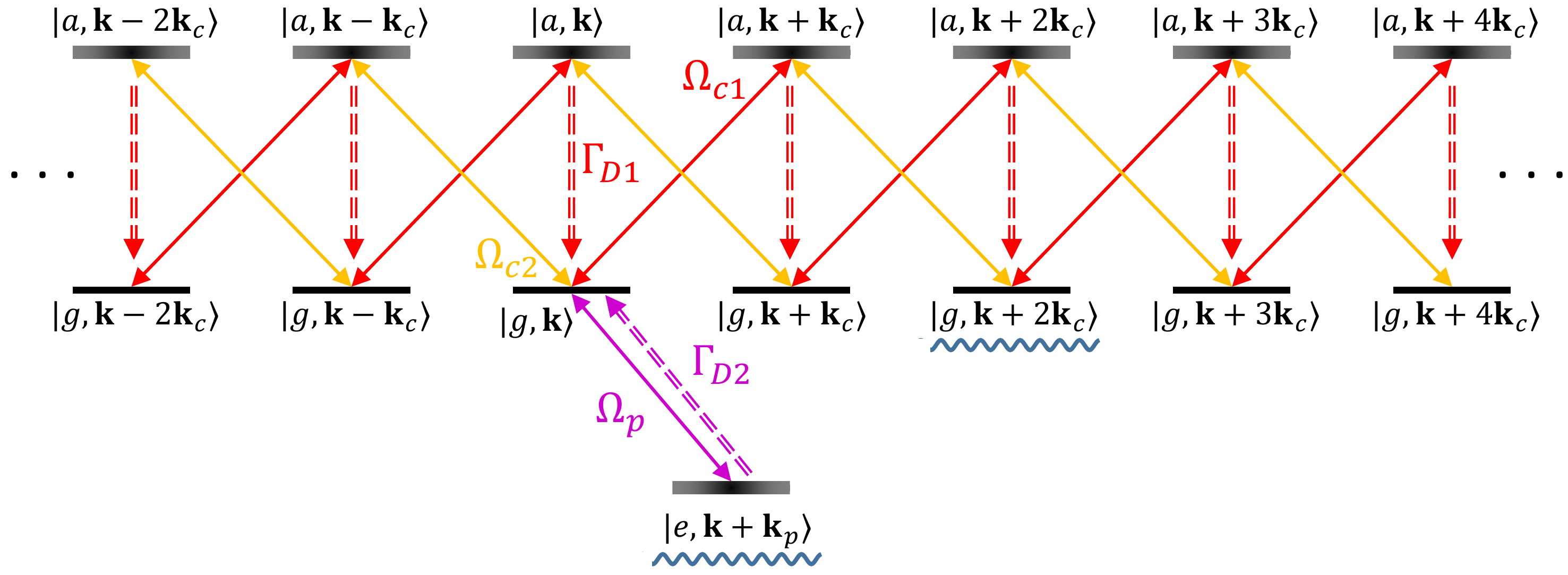} \caption{Momentum lattice structure for probe excitation and $U_g$ control simulations according to Eqs.~(\ref{equHeff3})-(\ref{equiC}). Dash arrows represent the effective quantum jump operations associated with Eq.~(\ref{equiC}). The double-sided arrows represent the coherent laser couplings. The coherence between the wavy underlined lattice sites $|e,{\bf k}+{\bf k}_p\rangle$ and $|g,{\bf k}+2{\bf k}_c\rangle$ is associated with the redirected superradiant emission.} \label{figMLattice}
\end{figure}

\begin{figure} 
\includegraphics[width=0.4\textwidth]{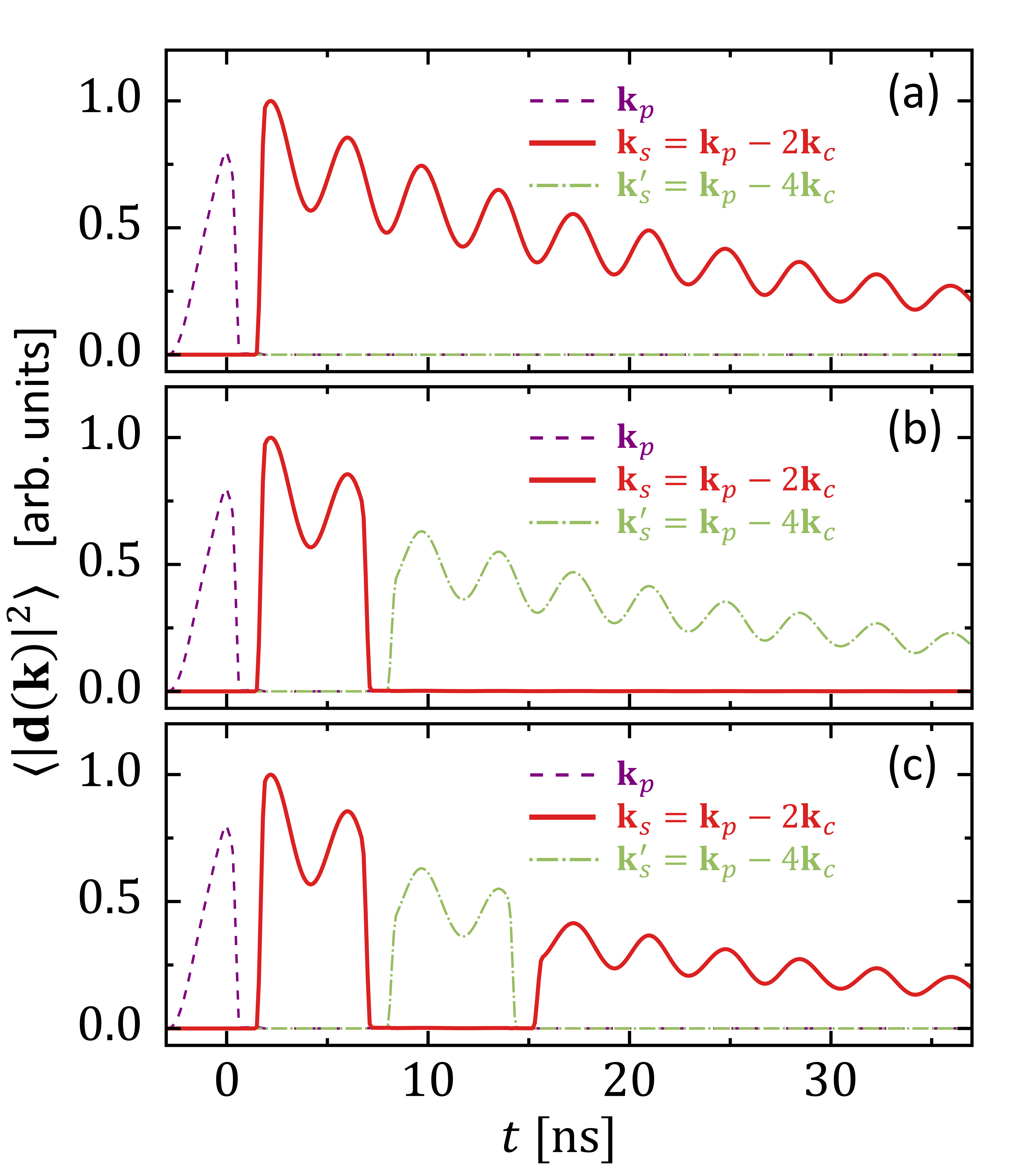} 
\caption{Simulation of spin-wave dynamics for typical experimental sequences. The spin-wave intensities represented by $\langle |{\bf d}({\bf k},t)|^2\rangle$ are evaluated according to Eq.~(\ref{equiMasterS}) with optimally estimated experimental parameters. (a) is according to Eq.~(\ref{equHeff3}) with a $\tau_p=3$~ns $D2$ probe excitation followed by a ${\bf k}\rightarrow {\bf k}-2{\bf k}_c$ control composed of two chirped $D1$ pulses with $\tau_c=0.5$~ns and $\tau_d=1.24$~ns at $\Delta t_1=0.2$~ns. The ${\bf k}_p$ (corresponds to forward radiation) and ${\bf k}_s={\bf k}_p-2{\bf k}_c$ (corresponds to the redirected radiation) spin-wave components are plotted with dashed and solid lines, respectively. The dash-dotted line corresponds to the mismatched ${\bf k}_s'={\bf k}_p-4{\bf k}_c$ excitation. In (b), an additional ${\bf k}\rightarrow {\bf k}-2{\bf k}_c$ is applied with $\Delta t_2=4.8$~ns. In (c), a ${\bf k}\rightarrow {\bf k}+2{\bf k}_c$ recall is simulated at $\Delta t_3=5.5$ ns. A recall efficiency similar to those in Fig.~\ref{figsample} is recovered. The $\sim 267~$MHz quantum beat amplitude needs to be taken into account when comparing the spin wave amplitudes
and to estimate the control efficiency. Notice the difference of $\tau_p$ and thus a different contrast to the quantum beat comparing with measurements in Fig.~\ref{figsample}. } \label{simulatefigureonoff}
\end{figure}

\subsection{$f_d$ and $f_a$ estimation}

We relate the experimental observable $i_{{\bf k}_s}(t)$ with ensemble-averaged $i_{s,\eta}(t)$ as $\langle i_{s,\eta}\rangle_{\eta}$, and calculate collective dipole control efficiency as $f_d=\langle i_{s,\eta}(\tau_c+\tau_d)\rangle_{\eta} /i_{s,0}(0)$. The ensemble average of emission intensity, instead of field amplitude, is in light of the fact that we experimentally collect $i_{{\bf k}_s}(t)$ with a multi-mode fiber, and the signal $i_{{\bf k}_s}(t)$ is insensitive to slight distortion of the ${\bf E}_s$-mode profile by the dynamic phase writing due to the imbalanced $\eta_{1,2}$.
  
The simulation of optical acceleration by the $D1$ control pulses follows the same Eqs.~(\ref{equHeff3}) and (\ref{equiMasterS}), but without the probe excitation and with atomic levels restricted to the $D1$ line only. We evaluate the momentum transfer as $\Delta P_{\eta} =\hbar {\bf k}_c\big(\sum_{g,n} n \langle g,{\bf k}+ n {\bf k}_c|\rho_{\eta}^{(s)}(t)|g, {\bf k}+ n {\bf k}_c\rangle+\sum_{a,n} n \langle a, {\bf k}+ n  {\bf k}_c|\rho_{\eta}^{(s)}(t)|a, {\bf k} + n {\bf k}_c\rangle\big)$ for $t=\tau_c+\tau_d$. We then compare the ensemble-averaged acceleration efficiency $f_a=\langle \Delta P_{\eta} \rangle _{\eta}/(2\hbar k_c)$ with the experimental measurements. 
  
The $\eta_{1,2}$ average in both calculations is according to spatial distribution of the control laser beam intensity profile seen by the atomic sample. As the final results are quite insensitive to distribution details, we assume both the laser beam and the atomic sample have Gaussian profiles, with waists $w=13~\mu$m and $\sigma=7~\mu$m by fitting the imaging measurements and with optics simulations. We adjust the retro-reflected beam waist $w_r$ and the intensity factor $\eta_2\propto 1/w_r$ accordingly in the simulation, together with an overall intensity calibration factor $\kappa$ multiplied to the $s$ parameter from the beat-note measurements of the control pulses~\cite{coSub}. The ensemble-averaged $f_a$ is compared with experimentally measured $ \Delta P/2\hbar k_c$, and we adjust $\kappa,w_r$ to globally match the single-atom simulation with all the measurement results for optical acceleration as in Fig.~\ref{figAcc}. We then estimate both $f_a$,$f_d$ as discussed in Sec.~\ref{secefficiency}. 
Since the second type of delay line~[Fig.~\ref{figTwodelayline}(b)] involves more optics than the first one~[Fig.~\ref{figTwodelayline}(a)], we expect more power loss and wavefront distortion for the retro-reflected pulses. However, with the second-type delay line, we are able to experimentally adjust the amplitude of the preprogrammed pulses [marked with ``$p'$'' in~Fig.~\ref{figTwodelayline}(b)] to approximately re-balance the intensity of the counter-propagating control pulses despite the power loss. In the corresponding simulation, we accordingly readjust the beam waist $w_r$ for the retro-reflection together with the intensity factor $\eta_2 \propto \alpha /w_r$. Within reasonable adjustments, the simulation results (see Fig.~\ref{simulatefigureonoff}) suggest that the efficiency $f_d$ of the single ${\bf k}$-shift operation reaches 75\%, agreeing with the estimation based on the retrieval efficiency~\cite{coSub}.


\subsection{Reconstructing the controlled spin-wave dynamics}

With the simulation parameters optimally matching the experiment, we further simulate the experimental sequence in Fig.~\ref{figsample} and calculate $\langle |{\bf d}({\bf k})|^2\rangle=\langle |{\rm tr}\big(\rho_{\eta}^{(s)}(t){\bf d}({\bf k})\big)|^2\rangle_{\eta}$ associated with collective dipole excitation with ${\bf k}=\{{\bf k}_p,{\bf k}_s={\bf k}_p-2{\bf k}_c,{\bf k}'_s={\bf k}_p-4{\bf k}_c\}$ for the forward, redirected, and subradiantly stored collective radiation, respectively.  We not only reproduce features of the experimental observable $i_{{\bf k}_s}(t)\propto \langle |{\bf d}({\bf k}_s)|^2\rangle$, but also unveil time-dependent dynamics for the unmonitored forward emission $\langle |{\bf d}({\bf k}_p)|^2\rangle$ and the subradiantly stored or the superradiance-free excitation $\langle |{\bf d}({\bf k}_s')|^2\rangle$. Typical results are given in Fig.~\ref{simulatefigureonoff}.

\subsection{Consistent recall at selected $\Delta t_3$ delay}

\begin{figure} 
    \includegraphics[width=0.4\textwidth]{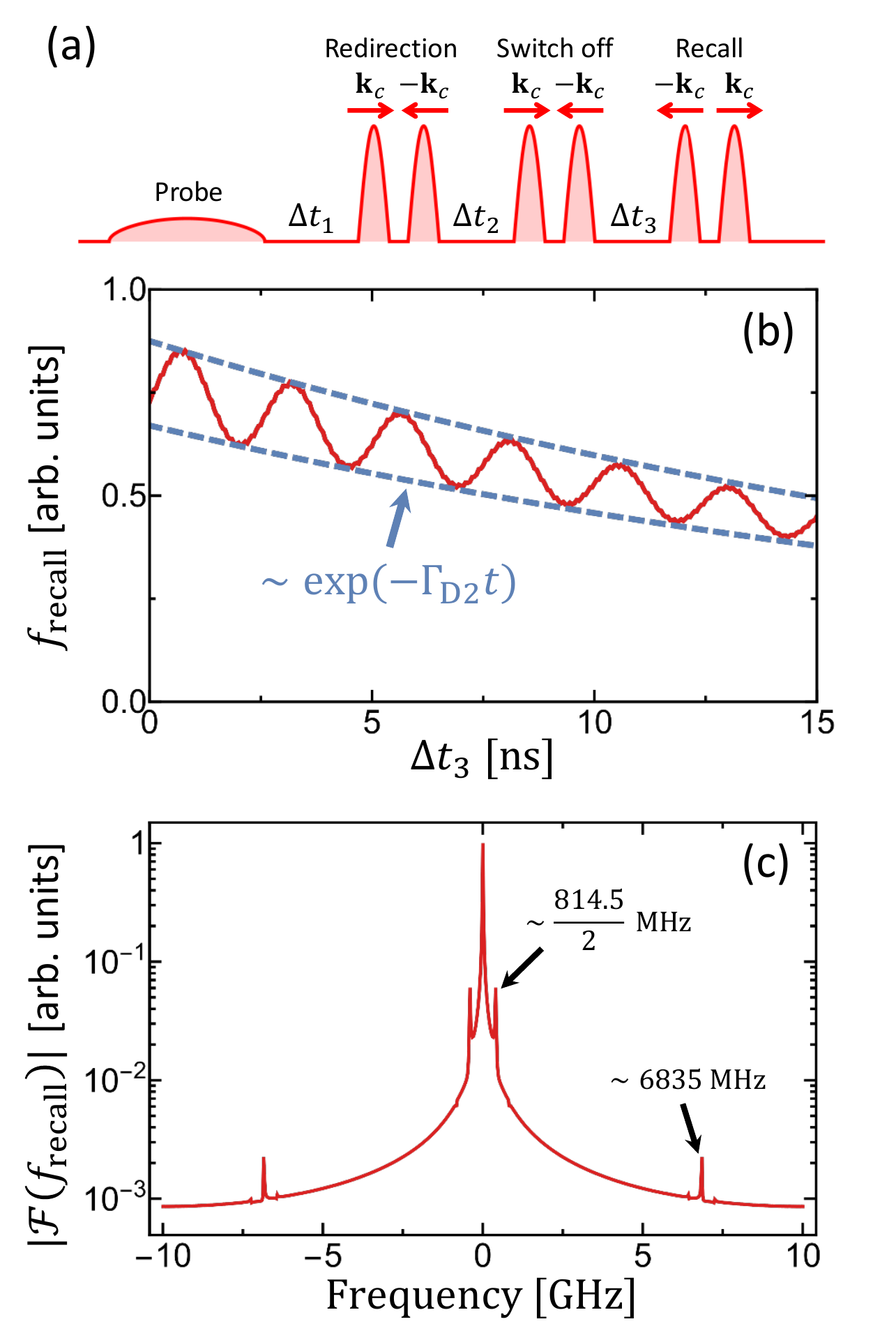} 
    \caption{Simulation of residual $D1$ dipole coherence induced interference effect for the recall with various $\Delta t_3$. (a) The pulse sequence of the simulation, according to the experiments. (b) We scan $\Delta t_3$ with all the other parameters in the full redirection--switch off--recall sequence unchanged and numerically calculate the relative recall efficiency. Blue dashed line: Reference of single decay dynamics. (c) The Fourier transform of the $\Delta t_3$-dependent recall efficiency.} \label{figrecallf}
\end{figure}

The redirection--switch off--recall sequence as in Fig.~\ref{figsample} involves multiple spin-wave controls $U_g(\pm 2{\bf k}_c)$ implemented with cyclic $D1$ transition, driven by pairs of counter-propagating control pulses. The controls are not perfect. In particular, after each pair of pulsed control, there is residual $D1$ population left in the 5$P_{1/2}$ state with dipole coherence determined by the optical phases as well as details of the control dynamics.  For two successive pulses of the same type, the coherently added residual $D1$ coherences may be enhanced or canceled, depending on the relative phase between the two. Such interference effects are actually explored for robust quantum control with composite pulse techniques~\cite{Genov2014, Chuang2016}. 

In this experiment, the interference effect emerges, in particular, between the second pulse of the switch-off control and the first pulse of the recall control, both driven by a $-{\bf k}_c$ chirped pulse [Fig.~\ref{figrecallf}(a)]. The relative phase between the dipole coherence evolves according to the laser detuning to the dipole transitions.  Here, with the center frequency of the control pulse at the midpoint of the hyperfine transitions 5$S_{1/2}~F=2$ -- 5$P_{1/2}~F'=1,2$, the interference leads to oscillation of the control efficiency, which is expected at the frequency $\Delta_{D1, {\rm hfse}} / 4 \pi = 814.5/2$~MHz ($\Delta_{D1, {\rm hfse}}$ is the 5$P_{1/2}$ hyperfine splitting.). The oscillations have been observed experimentally. To confirm the picture, we did a single-body simulation including all hyperfine levels in the $D1$ and $D2$ line of $^{87}{\rm Rb}$ as modeled in Appendix~\ref{SecExpModel}, by setting the Hamiltonian $H_{c1}+H_{c2}$ to match the sequence in the Fig.~\ref{figTimingsequence}(a) of the main text and furthermore with control parameters consistent with the experimental condition. All the parameters are fixed except the ``recall'' delay $\Delta t_3$, which is scanned in simulations to  numerically evaluate a relative ``recall'' efficiency define as $ f_{\rm recall} \propto \big( |{\bf d}({\bf k}_s)|^2 \big)\big|_{t=3\tau_c+3\tau_d+\Delta t_1 + \Delta t_2 +\Delta t_3} $.
The simulation results are shown in Fig.~\ref{figrecallf}. As expected, the ``recall'' efficiency oscillates with $\Delta t_3$ with $\Delta_{D1, {\rm hfse}} /2$ frequency.  In addition, it should be noted that there is oscillation at the frequency $\sim 6.8$~GHz with much smaller amplitude [Fig.~\ref{figrecallf}(b)], which is associated with the hyperfine splitting of the ground state. This additional modulation of the recall efficiency is because that the dipole coherence in the hyperfine transitions 5$S_{1/2}~F=1$ -- 5$P_{1/2}~F'=1,2$ is also excited during the control, though inefficiently since the control pulse is far-detuned, and negligibly influences our measurements. 

To make sure that the residual dipole coherence interference effect is consistent for all recall delays, we carefully select the recall delay time to be $\Delta t_3 = 4 \pi m / \Delta_{D1, {\rm hfse}} + t_{\rm off} $, with integers $m$ and a constant offset $t_{\rm off}$. This recall time selection makes it possible for us to study the decay dynamics of the phase-mismatched spin waves with a suppressed systematic error induced by the imperfect $D1$ control.


\begin{thebibliography}{99}




\bibitem{Quantuminformationandcomputation}
C.~H.~Bennett and D.~P.~Divincenzo,
\newblock Quantum information and computation,
\newblock  Nature (London) {\bf 404}, 247 (2000).
    
\bibitem{Colloquiumgaugepotential}
J.~Dalibard, F.~Gerbier, G.~Juzeli\={u}nas, and P.~\"Ohberg,
\newblock Colloquium: Artificial gauge potentials for neutral atoms,
\newblock  Rev. Mod. Phys. {\bf 83}, 1523 (2011).
    
\bibitem{quantumsimulation}
I.~M. Georgescu, S.~Ashhab, and F. Nori,
\newblock Quantum simulation,
\newblock  Rev. Mod. Phys. {\bf 86}, 153 (2014).
    
\bibitem{quantumsensing}
C.~L. Degen, F.~Reinhard, and P.~Cappellaro,
\newblock Quantum sensing,
\newblock  Rev. Mod. Phys. {\bf 89}, 035002 (2017).
    
\bibitem{quantummemoryreview}
K.~Hammerer, A.~S.~S\o{}rensen, and E.~S. Polzik,
\newblock Quantum interface between light and atomic ensembles,
\newblock  Rev. Mod. Phys. {\bf 82}, 1041 (2010).
    
\bibitem{quantummemoryreview2}
K.~Heshami, D.~G.~England, P.~C.~Humphreys, P.~J.~Bustard, V.~M.~Acosta, J.~Nunn, and B.~J.~Sussman,
\newblock Quantum memories: Emerging applications and recent advances,
\newblock  J. Mod. Opt. {\bf 63}, 2005 (2016).
    


\bibitem{Dicke1954}
R.~H. Dicke,
\newblock Coherence in spontaneous radiation processes,
\newblock  Phys. Rev. {\bf 93}, 99 (1954).


\bibitem{singlePhotondu2012}
S. Zhang, C. Liu, S. Zhou, C.-S. Chuu, M. M. T. Loy, and S. Du,
\newblock Coherent Control of Single-Photon Absorption and Reemission in a Two-Level Atomic Ensemble,
\newblock  Phys. Rev. Lett. {\bf 109}, 263601 (2012).

\bibitem{scully2006}
M.~O.~Scully, E.~S.~Fry, C.~H.~R.~Ooi, and K.~W\'{o}dkiewicz,
\newblock Directed Spontaneous Emission from an Extended Ensemble of $N$ Atoms: Timing Is Everything,
\newblock  Phys. Rev. Lett. {\bf 96}, 010501 (2006).

\bibitem{superradianceAraujo16}
M.~O.~Ara\'{u}jo, I.~Kre\v{s}i\'{c}, R.~Kaiser, and W.~Guerin,
\newblock Superradiance in a Large and Dilute Cloud of Cold Atoms in the Linear-Optics Regime,
\newblock  Phys. Rev. Lett. {\bf 117}, 073002 (2016).

\bibitem{superradianceRoof2016}
S.~J.~Roof, K.~J.~Kemp, M.~D.~Havey, and I.~M.~Sokolov,
\newblock Observation of Single-Photon Superradiance and the Cooperative Lamb Shift in an Extended Sample of Cold Atoms,
\newblock Phys. Rev. Lett. {\bf 117}, 073003 (2016).


\bibitem{Fleischhauer2005}
M.~Fleischhauer, A.~Imamoglu, and J.~P.~Marangos,
\newblock Electromagnetically induced transparency: Optics in coherent media,
\newblock Rev. Mod. Phys. {\bf 77}, 633 (2005).





\bibitem{ZhuPRA2016}
B.~Zhu, J.~Cooper, J.~Ye, and A.~M.~Rey,
\newblock Light scattering from dense cold atomic media,
\newblock  Phys. Rev. A {\bf 94}, 023612 (2016).

\bibitem{Chang2020RG}
F.~Andreoli, M.~J.~Gullans, A.~A.~High, A.~Browaeys, and D.~E.~Chang,
\newblock The maximum refractive index of an atomic medium,
\newblock  arXiv:2006.01680.





\bibitem{Jun2016}
S. L. Bromley, B. Zhu, M. Bishof, X. Zhang, T. Bothwell, J. Schachenmayer, T. L. Nicholson, R. Kaiser, S. F. Yelin, M. D. Lukin, A. M. Rey, and J. Ye,
\newblock Collective atomic scattering and motional effects in a dense coherent medium,
\newblock  Nat. Commun. {\bf 7}, 11039 (2016).

\bibitem{clScullyPRL09}
M.~O.~Scully,
\newblock Collective Lamb Shift in Single Photon Dicke Superradiance,
\newblock  Phys. Rev. Lett.  {\bf 102}, 143601 (2009).

\bibitem{Kaiser2016}
W. Guerin, M. T. Rouabah, and R. Kaiser,
\newblock Light interacting with atomic ensembles: Collective, cooperative and mesoscopic effects,
\newblock  J. Mod. Opt. {\bf 64}, 895 (2017).

\bibitem{Du2015}
S.-T. Chui, S. Du, and G.-B. Jo,
\newblock Subwavelength transportation of light with atomic resonances,
\newblock  Phys. Rev. A {\bf 92}, 053826 (2015).

\bibitem{Adams2016}
R.~J.~Bettles, S.~A.~Gardiner, and C.~S.~Adams,
\newblock Cooperative eigenmodes and scattering in one-dimensional atomic arrays,
\newblock  Phys. Rev. A {\bf 94}, 043844 (2016).

\bibitem{Needham2019}
J.~A.~Needham, I.~Lesanovsky, and B.~Olmos,
\newblock Subradiance-protected excitation transport,
\newblock  New J. Phys. {\bf 21}, 073061 (2019).

\bibitem{Bettles2016}
R.~J.~Bettles, S.~A.~Gardiner, and C.~S.~Adams,
\newblock Enhanced Optical Cross Section via Collective Coupling of Atomic Dipoles in a 2D Array,
\newblock Phys. Rev. Lett. {\bf 116}, 103602 (2016).

\bibitem{Shahmoon2017}
E. Shahmoon, D. S. Wild, M. D. Lukin, and S. F. Yelin,
\newblock Cooperative Resonances in Light Scattering from Two-Dimensional Atomic Arrays,
\newblock  Phys. Rev. Lett. {\bf 118}, 113601 (2017).




\bibitem{Immanuel2020}
J.~Rui, D.~Wei, A.~Rubio-Abadal, S.~Hollerith, J.~Zeiher, D.~M.~Stamper-Kurn, C.~Gross, and I.~Bloch,
\newblock A subradiant optical mirror formed by a single structured atomic layer,
\newblock Nature (London) {\bf 583}, 369 (2020).




\bibitem{selectiveRadiancePRXchang17}
A.~Asenjo-Garcia, M.~Moreno-Cardoner, A.~Albrecht, H.~J.~Kimble, and D.~E.~Chang,
\newblock Exponential Improvement in Photon Storage Fidelities Using Subradiance and ``Selective Radiance'' in Atomic Arrays,
\newblock  Phys. Rev. X {\bf 7}, 031024 (2017).

\bibitem{Sutherland2016}
R. T. Sutherland and F. Robicheaux,
\newblock Collective dipole-dipole interactions in an atomic array,
\newblock Phys. Rev. A {\bf 94}, 013847 (2016).

\bibitem{syznanov2016}
S.~V.~Syzranov, M.~L.~Wall, B.~Zhu, V.~Gurarie, and A.~M.~Rey,
\newblock Emergent Weyl excitations in systems of polar particles,
\newblock  Nat. Commun. {\bf 10}, 13543 (2016).

\bibitem{Perczel2017}
J.~Perczel, J.~Borregaard, D.~E. Chang, H.~Pichler, S.~F. Yelin, P.~Zoller, and M.~D.~Lukin,
\newblock Topological Quantum Optics in Two-Dimensional Atomic Arrays,
\newblock Phys. Rev. Lett. {\bf 119}, 023603 (2017).
 
\bibitem{Bettles2017}
R.~J.~Bettles, J.~Min\'{a}\v{r}, C.~S.~Adams, I.~Lesanovsky, and B.~Olmos,
\newblock Topological properties of a dense atomic lattice gas,
\newblock  Phys. Rev. A {\bf 96}, 041603(R) (2017).

\bibitem{Molmer2019}
Y.-X. Zhang and K. M\o{}lmer,
\newblock Theory of Subradiant States of a One-Dimensional Two-Level Atom Chain,
\newblock Phys. Rev. Lett. {\bf 122}, 203605 (2019).


\bibitem{coSub}
Y.~He, L.~Ji, Y.~Wang, L.~Qiu, J.~Zhao, Y.~Ma, X.~Huang, S.~Wu, and D.~E.~Chang,
\newblock Geometric Control of Collective Spontaneous Emission,
\newblock Phys. Rev. Lett. {\bf 125}, 213602 (2020).



\bibitem{accMetcalf2007}
X.~Miao, E.~Wertz, M.~G. Cohen, and H.~Metcalf,
\newblock Strong optical forces from adiabatic rapid passage,
\newblock  Phys. Rev. A {\bf 75}, 011402(R) (2007).


\bibitem{nonAdiabaticMetcalf2007}
T.~Lu, X.~Miao, and H.~Metcalf,
\newblock Nonadiabatic transitions in finite-time adiabatic rapid passage,
\newblock  Phys. Rev. A {\bf 75}, 063422 (2007).


\bibitem{MetcalfRevew2017}
H.~Metcalf,
\newblock Colloquium: Strong optical forces on atoms in multifrequency light,
\newblock  Rev. Mod. Phys. {\bf 89},  041001 (2017).

\bibitem{slowCampbell2014}
A. M. Jayich, A. C. Vutha, M. T. Hummon, J. V. Porto, and W. C. Campbell,
\newblock Continuous all-optical deceleration and single-photon cooling of molecular beams,
\newblock  Phys. Rev. A {\bf 89}, 023425 (2014).

\bibitem{campbell2019}
X.~Long, S.~S.~Yu, A.~M.~Jayich, and W.~C.~Campbell,
\newblock Suppressed Spontaneous Emission for Coherent Momentum Transfer,
\newblock  Phys. Rev. Lett. {\bf 123}, 033603 (2019).


\bibitem{DaWei2014superAI}
D.-W.~Wang and M.~O.~Scully,
\newblock Heisenberg Limit Superradiant Superresolving Metrology,
\newblock Phys. Rev. Lett. {\bf 113}, 083601 (2014).


\bibitem{scully2015}
M.~O.~Scully,
\newblock Single Photon Subradiance: Quantum Control of Spontaneous Emission and Ultrafast Readout,
\newblock  Phys. Rev. Lett. {\bf 115}, 243602 (2015).



\bibitem{Monroe2014}
J. Mizrahi, B. Neyenhuis, K. G. Johnson, W. C. Campbell, C. Senko, D. Hayes, and C. Monroe,
\newblock Quantum control of qubits and atomic motion using ultrafast laser pulses,
\newblock  Appl. Phys. B {\bf 114}, 45 (2014).

\bibitem{Monroe2017}
J. D. Wong-Campos, S. A. Moses, K. G. Johnson, and C. Monroe,
\newblock Demonstration of Two-Atom Entanglement with Ultrafast Optical Pulses,
\newblock Phys. Rev. Lett. {\bf 119}, 230501 (2017).

\bibitem{houge2018}
M.~Jaffe, V.~Xu, P.~Haslinger, H.~M\"uller, and P.~Hamilton,
\newblock Efficient Adiabatic Spin-Dependent Kicks in an Atom Interferometer,
\newblock  Phys. Rev. Lett. {\bf 121}, 040402 (2018).

\bibitem{cundiff2013}
S.~T.~Cundiff and S.~Mukamel, 
\newblock Optical multidimensional coherent spectroscopy,
\newblock  Phys. Today {\bf 66}, 44 (2013).

\bibitem{Fuller2015}
F.~D.~Fuller and J.~P.~Ogilvie, 
\newblock Experimental implementations of two-dimensional Fourier transform electronic spectroscopy,
\newblock  Annu. Rev. Phys. Chem. {\bf 66}, 667 (2015).

\bibitem{Oliver2018}
T.~A.~A.~Oliver, 
\newblock Recent advances in multidimensional ultrafast spectroscopy,
\newblock  R. Soc. Open Sci. {\bf 5}, 171425 (2018).

\bibitem{geometricAspect2012}
T. Ichikawa, M. Bando, Y. Kondo, and M. Nakahara,
\newblock Geometric aspects of composite pulses,
\newblock Phil. Trans. R. Soc. A {\bf 370}, 4671 (2012).

\bibitem{Loy1974}
M. M. T. Loy,
\newblock Observation of Population Inversion by Optical Adiabatic Rapid Passage,
\newblock  Phys. Rev. Lett. {\bf 32}, 814 (1974).



\bibitem{Shapiro2008}
S. Zhdanovich, E. A. Shapiro, M. Shapiro, J. W. Hepburn, and V. Milner,
\newblock Population Transfer between Two Quantum States by Piecewise Chirping of Femtosecond Pulses: Theory and Experiment,
\newblock Phys. Rev. Lett. {\bf 100}, 103004 (2008).


\bibitem{GoswamiPR2003}
D. Goswami,
\newblock Optical pulse shaping approaches to coherent control,
\newblock Phys. Rep. {\bf 374}, 385 (2003).

\bibitem{scullyscience09}
M.~O.~Scully and A.~A.~Svidzinsky, 
 \newblock The super of superradiance,
\newblock Science {\bf 325}, 1510 (2009).


\bibitem{shortPaper2}
The spin-wave control technique combined with modified trap geometries allows us to precisely characterize the phase-(mis)matched dipole spin waves dynamics as a function of atomic density and number. A paper to present the findings is under preparations.

\bibitem{DEC2019PRA}
L.~Henriet, J.~S.~Douglas, D.~E.~Chang, and A.~Albrecht,
 \newblock Critical open-system dynamics in a one-dimensional optical-lattice clock,
\newblock  Phys. Rev. A {\bf 99}, 023802 (2019).

\bibitem{Steck2015}
D.~A.~Steck, Rubidium 87 D Line Data, http://steck.us/alkalidata (revision 2.2.1, November 21, 2019).




\bibitem{superradianceCottier18}
F.~Cottier, R.~Kaiser, and R.~Bachelard,
\newblock Role of disorder in super- and subradiance of cold atomic clouds,
\newblock Phys. Rev. A {\bf 98}, 013622 (2018).


\bibitem{foot:probenormalization}
The data in Fig.~\ref{figDensity}(b) is from automated measurements over 12 
hours with three rounds of automation, with probe power being increased by 
up to a factor of 5 to enhance the signals for  measurements with smallest 
atom numbers. The adjustments are confirmed not to affect the decay 
dynamics, and the probe intensities are all well within the linear 
excitation regime. We recorded the probe light power for each automation. 
The $i_{{\rm max},N}$ in Fig.~\ref{figDensity}(b) are relatively normalized by the 
recorded probe power ratio for the three sets of automation.


\bibitem{geometricCheneau2008}
M. Cheneau, S. P. Rath, T. Yefsah, K. J. G\"unter, G. Juzeli\=unas, and J. Dalibard,
\newblock Geometric potentials in quantum optics: A semi-classical interpretation,
\newblock  EPL {\bf 83}, 60001 (2008).

\bibitem{Gritsev2012}
V.~Gritsev and A.~Polkovnikov,
 \newblock Dynamical quantum Hall effect in the parameter space,
\newblock Proc. Natl. Acad. Sci. U.S.A. {\bf 109}, 6457 (2012).











\bibitem{foot:longlived}
For few-atoms system with long-lived excitations, techniques are developed 
for frequency-selective excitation of the collective modes, for example, by 
Z. Meir {\it et al.} in  Phys. Rev. Lett. {\bf 113}, 193002 (2014); and by B. H. 
McGuyer {\it et al.} in  Nat. Phys. {\bf 11}, 32 (2015). Time-domain control 
techniques are also developed to manipulate the collective modes, for 
example, by S.~de~L\'es\'eleuc {\it et al.} in  Phys. Rev. Lett. {\bf 119}, 
053202 (2017).

\bibitem{foot:superLattice}
As the collective states of many-atom system are typically with unresolved 
energy differences, precise control of isolated superradiant states in the 
frequency domain is generally difficult to achieve. However, dynamics of the 
collective states controlled by continuous-wave lasers with constant lattice 
couplings is an interesting and emergent topic that has intriguing consequences. For example,  superradiance lattices consisting of 
collective states have been discussed by D.-W.~Wang {\it et al.} in  Phys. 
Rev. Lett. {\bf 114}, 043602 (2015);
experimentally studied by  L.~Chen {\it et al.} in  Phys. Rev. Lett. {\bf 120}, 193601 (2018); and by H.~Cai {\it et al.} in  Phys. Rev. Lett. {\bf 122}, 023601 (2019), in both cold and hot atomic systems.

\bibitem{foot:otherField}
Beyond atomic physics, techniques are developed for super- and subradiant 
states conversion in other fields, such as by Z.~Wang {\it et al.} in Phys. Rev. Lett. {\bf 124}, 013601 (2020).

\bibitem{StAChen2010}
X. Chen, A. Ruschhaupt, S. Schmidt, A. del Campo, D.~Gu\'ery-Odelin, and 
J.~G.~Muga,
 \newblock Fast Optimal Frictionless Atom Cooling in Harmonic Traps: Shortcut to Adiabaticity,
\newblock  Phys. Rev. Lett. {\bf 104}, 063002 (2010).

\bibitem{freegarde2014}
A.~Dunning, R.~Gregory, J.~Bateman, N.~Cooper, M.~Himsworth, J.~A.~Jones, and T.~Freegarde,
 \newblock Composite pulses for interferometry in a thermal cold atom cloud,
\newblock Phys. Rev. A {\bf 90}, 033608 (2014).

\bibitem{NatCommDu2015}
X. Rong, J. Geng, F. Shi, Y. Liu, K. Xu, W. Ma, F. Kong, Z. Jiang, Y. Wu, and J. Du,
\newblock Experimental fault-tolerant universal quantum gates with solid-state spins under ambient conditions,
\newblock  Nat. Commun. {\bf 6}, 8748 (2015).

\bibitem{Chuang2016}
G.~H.~Low, T.~J.~Yoder, and I.~L.~Chuang,
\newblock Methodology of Resonant Equiangular Composite Quantum Gates,
\newblock  Phys. Rev. X {\bf 6}, 041067 (2016).

\bibitem{multiphotonMOT}
S.~Wu, T.~Plisson, R.~C.~Brown, W.~D.~Phillips, and J.~V.~Porto,
\newblock Multiphoton Magnetooptical Trap,
\newblock  Phys. Rev. Lett. {\bf 103}, 173003 (2009).


\bibitem{Yudi2020}
Y.~Ma, X.~Huang, X.~Wang, L.~Ji, Y.~He, L.~Qiu, J.~Zhao, Y.~Wang, and S.~Wu,
\newblock Precise pulse shaping for quantum control of strong optical transitions,
\newblock Opt. Express {\bf 28}, 17171 (2020).


\bibitem{Freegarde1995}
T.~G.~M.~Freegarde, J.~Walz, and T.~W.~H\"ansch,
\newblock Confinement and manipulation of atoms using short laser pulses,
\newblock  Opt. Commun. {\bf 117}, 262 (1995).

\bibitem{Immanuel1997}
A.~Goepfert, I.~Bloch, D.~Haubrich, F.~Lison, R.~Sch\"utze, R.~Wynands, and D.~Meschede,
 \newblock Stimulated focusing and deflection of an atomic beam using picosecond laser pulses,
\newblock  Phys. Rev. A {\bf 56}, R3354(R) (1997).

\bibitem{Monroe2007}
M. J. Madsen, D. L. Moehring, P. Maunz, R. N. Kohn, Jr., L.-M. Duan, and C. Monroe,
\newblock Ultrafast Coherent Excitation of a Trapped Ion Qubit for Fast Gates and Photon Frequency Qubits,
\newblock  Phys. Rev. Lett. {\bf 97}, 040505 (2006).

\bibitem{shaperTutorial2010}
A.~Monmayrant, S.~Weber, and B.~Chatel,
\newblock A newcomer's guide to ultrashort pulse shaping and characterization,
\newblock J. Phys. B {\bf 43}, 103001 (2010).

\bibitem{Weiner2011Review}
A.~M.~Weiner,
\newblock Ultrafast optical pulse shaping: A tutorial review,
\newblock Opt. Commun. {\bf 284}, 3669 (2011).



\bibitem{NMRAdiabatic2001}
M.~Garwood and L.~DelaBarre,
 \newblock The return of the frequency sweep: Designing adiabatic pulses for contemporary NMR,
\newblock   J. Magn. Reson. {\bf 153}, 155 (2001).

\bibitem{prl2016}
G.~Facchinetti, S.~D. Jenkins, and J.~Ruostekoski,
 \newblock Storing Light with Subradiant Correlations in Arrays of Atoms,
\newblock  Phys. Rev. Lett. {\bf 117}, 243601 (2016).

\bibitem{Corman2017}
L.~Corman, J.~L.~Ville, R.~Saint-Jalm,  M.~Aidelsburger, T.~Bienaim\'{e}, 
S.~Nascimb\`ene, J.~Dalibard, and J.~Beugnon,
 \newblock Transmission of near-resonant light through a dense slab of cold atoms,
\newblock Phys. Rev. A  {\bf 96}, 053629 (2017).


\bibitem{Jennewein2016}
S. Jennewein, M. Besbes, N. J. Schilder, S. D. Jenkins, C. Sauvan, J. Ruostekoski, J.-J. Greffet, Y. R. P. Sortais, and A. Browaeys,
 \newblock Coherent Scattering of Near-Resonant Light by a Dense Microscopic Cold Atomic Cloud,
\newblock  Phys. Rev. Lett. {\bf 116}, 233601 (2016).

\bibitem{Guimond2019}
P.-O.~Guimond, A.~Grankin, D.~V. Vasilyev, B.~Vermersch, and P.~Zoller,
\newblock Subradiant Bell States in Distant Atomic Arrays,
\newblock  Phys. Rev. Lett. {\bf 122}, 093601 (2019).

\bibitem{oeJavanainen2017}
J. Javanainen and J. Ruostekoski,
 \newblock Light propagation beyond the mean-field theory of standard optics,
\newblock  Opt. Express {\bf 24}, 993 (2016).


\bibitem{CarusottoRMP2013Rev}
 I. Carusotto and C. Ciuti,
   \newblock Quantum fluids of light,
\newblock Rev. Mod. Phys. {\bf 85}, 299 (2013).
  
\bibitem{DarrickNP2014Rev}
D.~E.~Chang, V.~Vuleti\'c, and M.~D.~Lukin,
\newblock Quantum nonlinear optics - photon by photon,
\newblock Nat. Photonics {\bf 8}, 685 (2014).
  
\bibitem{ChangsukRPP2017Rev}
C.~Noh and D.~G.~Angelakis,
\newblock Quantum simulations and many-body physics with light,
\newblock Rep. Prog. Phys. {\bf 80}, 016401 (2016).

\bibitem{ReinaudiOptLett2007}
G.~Reinaudi, T.~Lahaye, Z.~Wang, and D.~Gu\'ery-Odelin,
 \newblock Strong saturation absorption imaging of dense clouds of ultracold  atoms,
\newblock Opt. Lett. {\bf 32}, 3143 (2007).

\bibitem{Chin2017}
K. Hueck, N. Luick, L. Sobirey, T. Lompe, H. Moritz, L.
Clark, and C. Chin,
 \newblock Calibrating high intensity absorption imaging of ultracold atoms,
\newblock Opt. Express {\bf 25}, 8670 (2017).


\bibitem{Gross1982}
M. Gross and S. Haroche,
\newblock   Superradiance: An essay on the theory of collective spontaneous emission,
\newblock  Phys. Rep. {\bf 93}, 301 (1982).




\bibitem{Manassah12}
J. T. Manassah,
\newblock Cooperative radiation from atoms in different geometries: Decay rate and frequency shift,
\newblock Adv. Opt. Photon. {\bf 4}, 108 (2012).

\bibitem{subradianceGuerin2016}
W. Guerin, M.~O.~Ara\'ujo, and R. Kaiser,
\newblock Subradiance in a Large Cloud of Cold Atoms,
\newblock  Phys. Rev. Lett. {\bf 116}, 083601 (2016).

\bibitem{subRadianceKaiser2012}
T.~Bienaim\'{e}, N~Piovella, and R.~Kaiser,
 \newblock Controlled Dicke Subradiance from a Large Cloud of Two-Level Systems,
\newblock Phys. Rev. Lett. {\bf 108}, 123602 (2012).

\bibitem{Sutherland2016B}
R.~T.~Sutherland and F.~Robicheaux,
\newblock Coherent forward broadening in cold atom clouds,
\newblock Phys. Rev. A {\bf 93}, 023407 (2016).

\bibitem{Genov2014}
G. T. Genov, D. Schraft, T. Halfmann, and N. V. Vitanov,
 \newblock Correction of Arbitrary Field Errors in Population Inversion of Quantum Systems by Universal Composite Pulses,
\newblock  Phys. Rev. Lett. {\bf 113}, 043001 (2014).



\end{thebibliography}
\end{document}